\documentclass[twoside,english,nofootinbib,preprintnumbers,aps,11pt]{revtex4}

\usepackage{euscript,amssymb}
\usepackage{amsthm}
\usepackage{graphicx}
\usepackage{amsfonts}
\usepackage{amsmath}
\usepackage{amssymb}
\usepackage{fancyhdr}
\usepackage{epsfig}
\usepackage{color}
\usepackage{esint}
\usepackage{soul}
\usepackage{afterpage}
\usepackage[papersize={8.5in,11in}]{geometry}

\newcommand{\be}{\begin{equation}}
\newcommand{\ee}{\end{equation}}
\newcommand{\bra}[1]{\langle #1 |}
\newcommand{\ket}[1]{|#1\rangle}
\newcommand{\braket}[2]{\langle #1 | #2 \rangle}
\newcommand{\half}{\tfrac{1}{2}}

\newcommand{\makeSymbol}[1]{\mathord{\vcenter{\hbox{#1}}}}

\newcommand{\ppr}{\vec p^{\,\prime}}

\begin{document}

\title{Coherent State Operators in Loop Quantum Gravity}

\author{Emanuele Alesci}
\email[]{emanuele.alesci@fuw.edu.pl}
\author{Andrea Dapor}
\email[]{adapor@fuw.edu.pl}
\author{Jerzy Lewandowski}
\email[]{jerzy.lewandowski@fuw.edu.pl}
\author{Ilkka M\"{a}kinen}
\email[]{ilkka.makinen@fuw.edu.pl}
\author{Jan Sikorski}
\email[]{jan.sikorski@fuw.edu.pl}

\affiliation{\vspace{6pt} Instytut Fizyki Teoretycznej, Uniwersytet Warszawski, ul. Pasteura 5, 02-093 Warsaw, Poland}
\begin{abstract}

\noindent We present a new method for constructing operators in loop quantum gravity. The construction is an application of the general idea of ``coherent state quantization'', which allows one to associate a unique quantum operator to every function on a classical phase space. Using the heat kernel coherent states of Hall and Thiemann, we show how to construct operators corresponding to functions depending on holonomies and fluxes associated to a fixed graph. We construct the coherent state versions of the fundamental holonomy and flux operators, as well as the basic geometric operators of area, angle and volume. Our calculations show that the corresponding canonical operators are recovered from the coherent state operators in the limit of large spins.

\end{abstract}

\date{\today}

\pacs{???}

\maketitle
\section{Introduction}
Coherent states are used in virtually every area of quantum physics, due to their importance in relation to the classical limit of quantum theories \cite{CS1, CS2, CS3}. A minimum requirement for a family of coherent states, independent of the system studied, is that each state is labelled by a point in the phase space of the theory: the interpretation is that such quantum state represents the closest approximation to the classical state corresponding to the phase space point. Loop quantum gravity (LQG) \cite{LQG1, LQG2, LQG3} is not an exception to this principle: various proposals exist to represent (discrete) classical geometries in terms of quantum states in the kinematical Hilbert space of the theory. 

The basic idea behind these constructions comes from the weave states introduced in \cite{Ashtekar:1992tm}, where the authors built classical geometries using quantum states based on fixed graph structures dense enough to reproduce classical values for the intrinsic 3-geometries. Those states, using heat kernel techniques for compact groups \cite{Hall}, were generalized \cite{LQG-CS1, LQG-CS2, LQG-CS3, LQG-CS4, Thiemann:2002vj, Sahlmann:2001nv} and extensively studied \cite{CSdevelop, LQG-CS5, LQG-CS6,LQG-CSspinfoam}. See also \cite{Oriti1} for similar states with improved peakedness properties. 

However, the search for ``good coherent states" is still far from complete in LQG, the main issue being that the known proposals are limited to a fixed graph: the coherent states usually considered in LQG cannot be projected to cylindrical functions in the complete Hilbert space. If we naively take a linear combination of such fixed-graph coherent states on the label-set of graphs, we soon discover that there is no damping factor fast enough to make the norm of the state finite. This problem is of course due to the non-separability of the kinematical Hilbert space, and might be solved at the physical level. However, a definition of coherent states even at the diffeomorphism invariant level is still missing, though proposals exist \cite{LQG-CS7} and preliminary studies on collective variables appeared \cite{Oriti2}, and a new program to deal with coherent states in the context of a Born-Oppenheimer approximation has been settled \cite{CS-quant5, CS-quant6, CS-quant7}. For most present purposes, however, the fixed-graph coherent states seem to work fine. In particular, such coherent states have been shown to be peaked on areas and volumes corresponding to those of classical polyhedra \cite{LQG-CS2, LQG-CS3, LQG-CSspinfoam}, at least in the large-$j$ limit.

In this work we want to explore a different application of coherent states, which does not concern their usage as quantum states representing semiclassical geometries, but rather the construction of new operators on the kinematical Hilbert space of LQG. More specifically, given a certain graph, one can generalize the resolution of identity satisfied by coherent states to define operators on the Hilbert space of to such graph. This construction of operators via coherent states has been already considered in the context of quantum mechanics \cite{CS-quant1, CS-quant2} and quantum cosmology \cite{CS-quant3, CS-quant4}, and is usually referred to as ``coherent state quantization''. We think that this name might give rise to confusion, as it suggests that the quantization is different than the standard one. It should be instead underlined that the Hilbert space of the theory remains untouched: one simply has a different (and non-canonical) way of associating an operator to a given phase space function. For this reason, we prefer to the expression ``coherent states quantization'' the more proper ``coherent state operators''.

The reasons to consider coherent state operators as opposed to canonical operators (that is, operators obtained by writing the classical function in terms of fundamental variables and then ``putting the hats'') are many:
\begin{itemize}
\item The correspondence between phase space function and coherent state operator is unique (no ordering ambiguities).
\item The coherent state operator corresponding to a real positive function is automatically symmetric and positive-definite.
\item By construction, coherent state operators have a good semiclassical limit: let $A_f$ be the coherent state operator associated to phase space function $f$; then, taking the expectation value of $A_f$ on the coherent state peaked on phase space point $(q, p)$ produces $f(q, p)$ plus a correction of order $\hbar$.
\end{itemize}
There is a major difference between coherent state operators and canonical ones: coherent state operators in general do not represent the classical Poisson algebra. Indeed, if $A_q$ and $A_p$ denote the coherent state operators associated to the fundamental phase space variables (position $q$ and momentum $p$), then it is not true that $[A_q, A_p] = i\hbar$. While this might seem a drawback, we remind the reader that even the canonical operators realize the Poisson algebra only to some extent: in fact, it is not true in general that $[f(\hat{q}, \hat{p}), g(\hat{q}, \hat{p})] = i\hbar \widehat{\{f(q, p), g(q, p)\}}$. Given this observation, we do not regard as a weakness the fact that coherent state operators forget completely about Poisson algebra, in favor of a more unique construction and clearer semiclassical interpretation.

This work is a study of coherent state operators in the context of LQG, accounting for general properties and providing a systematic set of examples. Specifically, the structure of the paper is the following. In Section \ref{CS-on-su2} we recall the definition and properties of $SU(2)$ coherent states on a fixed graph. In Section \ref{CS-general} the general construction of coherent state operators is presented. This construction is specialized to LQG in Section \ref{CS-examples}, where the coherent state operators counterparts of holonomy, flux, angle, area and volume are constructed and studied. The ``new'' holonomy-flux algebra is explicitly computed, to show that indeed it does not coincide with the classical one. We conclude in Section \ref{conclu} with closing remarks and outlooks.
\section{Coherent states on $SU(2)$} \label{CS-on-su2}
Coherent states adapted to a fixed graph are realized starting from coherent states on a single copy of the group $SU(2)$. In this case, the configuration space is $SU(2)$ itself, and hence the phase space is given by $\Gamma = T^*SU(2) \simeq SU(2) \times \mathfrak{su}_2$. This space is isomorphic to $SL(2,\mathbb{C})$, and as such each phase space point can be represented by an element $h \in SL(2,\mathbb{C})$. There are two particularly natural representations for a given $h$, which we call the left- and the right-representation:
\begin{align} \label{h-gp-reps}
h = g e^{t\vec p\cdot\vec\sigma/2} \ \ \ \ \ \text{and} \ \ \ \ \ h = e^{t\ppr\cdot\vec\sigma/2} g
\end{align}
where $g \in SU(2)$ and $\xi = -it\vec p\cdot\vec\sigma/2$ is an element of $\mathfrak{su}_2$ (expressed in terms of $\vec{p} \in \mathbb{R}^3$ and Pauli matrices $\vec{\sigma}$), and $t$ is a parameter which will be related to the semiclassical properties of the coherent state. Note that the variables $\vec p$ and $\ppr$ defined by the two decompositions are related by
\be\label{pandp'}
\ppr\cdot\vec\sigma = g(\vec p\cdot\vec\sigma)g^{-1}.
\ee
The two equivalent expressions \eqref{h-gp-reps} make explicit the relation between $(g, \vec{p}) \in \Gamma$ and $h \in SL(2, \mathbb{C})$.
\\
\\
The Hilbert space of this system is
\begin{align}\label{HS-link}
\mathcal{H} = L_2(SU(2), d\mu_H)
\end{align}
where $d\mu_H$ is the Haar measure on $SU(2)$. A useful basis for this space is the spin-network basis, whose elements will be denoted by $| j, m, n \rangle$. Here, $j = 0, 1/2, 1, 3/2, ...$ labels the irreducible representations of $SU(2)$, while $m$ and $n$ are ``magnetic indices'' taking values from $-j$ to $+j$ in integer steps. As a function in $\mathcal{H}$, a spin-network is given by Wigner matrices:\footnote
{
In the paper we use only lower magnetic indices for typographical reasons. In particular, when we write $D^{(j)}_{mn}$, the index $m$ should really be up. For this reason, we will be adopting summation convention over repeated magnetic indices, irrespective of their position.
}
\begin{align}
\langle g | j, m, n \rangle = \sqrt{d_j} \overline{D^{(j)}_{mn}(g)}
\end{align}
where $d_j = 2j + 1$ is the dimension of the $j$-irrep and the bar means complex conjugation. In this Hilbert space we have a representation of the fundamental operators corresponding to the classical phase space variables:
\begin{itemize}
\item The holonomy operator, which acts by multiplication:
\begin{align}
\left(\hat{D}^{(j)}_{mn} \Psi\right)(g) = {D}^{(j)}_{mn}(g) \Psi(g)
\end{align}
\item The flux operator, which acts by left-invariant or right-invariant derivation:
\begin{align}
\left(\hat{L}^i \Psi\right)(g) = \lim_{\epsilon \rightarrow 0} \dfrac{d}{d\epsilon} \Psi(g e^{-i\epsilon\sigma^i/2}) \ \ \ \ \ \ \ \ \ \ \left(\hat{R}^i \Psi\right)(g) = \lim_{\epsilon \rightarrow 0} \dfrac{d}{d\epsilon} \Psi(e^{i\epsilon\sigma^i/2} g)
\end{align}
\end{itemize}
In particular, on the spin-network basis these operators have the following matrix elements:
\begin{align} \label{standard-holo-me}
\langle j_1, m_1, n_1 | \hat{D}^{(j)}_{mn} | j_2, m_2, n_2 \rangle = \sqrt{\dfrac{d_{j_1}}{d_{j_2}}} {C^{j_1jj_2}_{m_1m m_2}} C^{j_1 jj_2}_{n_1 n n_2},
\end{align}
where $C^{j_1jj_2}_{m_1mm_2}$ denotes the Clebsch-Gordan coefficient (commonly also written as $\braket{j_1m_1jm}{j_2m_2}$), and
\begin{align} \label{standard-flux-me}
\langle j_1, m_1, n_1 | \hat{L}_i | j_2, m_2, n_2 \rangle & = \dfrac{i}{2} \delta_{j_1 j_2} \delta_{m_1 m_2} {}^{(j_1)}\sigma^i_{n_2n_1}, \\ \langle j_1, m_1, n_1 | \hat{R}_i | j_2, m_2, n_2 \rangle & = -\dfrac{i}{2} \delta_{j_1 j_2} {}^{(j_1)}\sigma^i_{m_2m_1}\delta_{n_1 n_2}, \label{standard-R-me}
\end{align}
where ${}^{(j)}\sigma^i_{mn}$ are the standard Hermitian generators of $SU(2)$ in the spin-$j$ representation.

%
%
%
		\subsection{Coherent states: Definition and semiclassical properties}
Each member of a coherent state family must be labeled by a point in phase space, $(g_0, \vec{p}_0) \in \Gamma$. The $SU(2)$ coherent states considered in this paper will thus be denoted $|g_0, \vec{p}_0\rangle$. The wave function of the state in the group representation is given by
\begin{align} \label{CS-in-g}
\psi^t_{(g_0, \vec{p}_0)}(g) \equiv \langle g | g_0, \vec{p}_0 \rangle = \sum_{j} d_j e^{-t  \lambda_j/2} \chi^{(j)}(g_0 e^{t\vec{p}_0 \cdot \vec{\sigma}/2}g^{-1})
\end{align}
where $\lambda_j := j(j + 1)$ and $\chi^{(j)}(g) := {\rm Tr}\,D^{(j)}(g)$, and $t$ parametrizes the spread of the state $\psi^t_{(g_0, \vec{p}_0)}$. In Eq. \eqref{CS-in-g} we have chosen to decompose $h$ in the left-representation $h = g_0 e^{t\vec{p}_0 \cdot \vec{\sigma}/2}$. The components of this state in the spin-network basis are
\begin{align} \label{CS-in-sn}
\widetilde\psi^t_{(g_0, \vec{p}_0)}(j,m,n) \equiv \langle j, m, n | g_0, \vec{p}_0 \rangle = \sqrt{d_j} e^{-t  \lambda_j/2} {D^{(j)}_{mn}(g_0 e^{t\vec{p}_0 \cdot \vec{\sigma}/2})}.
\end{align}
At this point one can prove various semiclassical properties of $\psi^t_{(g_0, \vec{p}_0)}$ (see \cite{LQG-CS1, LQG-CS2, LQG-CS3} for a complete account). In particular,
\begin{itemize}
\item $\bigl|\psi^t_{(g_0, \vec{p}_0)}(g)\bigr|^2$ is peaked on $g=g_0$. The width of the peak is controlled by the parameter $t$, and the peak becomes sharp when $t\ll 1$.
\item $\bigl|\widetilde\psi^t_{(g_0, \vec{p}_0)}(j,m,n)\bigr|^2$ is peaked on $j\simeq |\vec p_0|$, $m \simeq (p_0')_z$, $n \simeq (p_0)_z$, where $\vec p$ and $\ppr$ refer to the left- and right-decompositions of the $SL(2,\mathbb{C})$ element $h$. If the value of $t$ is fixed\footnote{Say, by the requirement that the coherent state is well peaked on the group element.}, then the peak in ''momentum space'' is sharp when $|\vec p_0| \gg 1/\sqrt t$.
\end{itemize}
These results justify the statement that the state $|g_0, \vec{p}_0\rangle$ is peaked on the classical phase space point $(g_0, \vec{p}_0)$, and for this reason is be said to be semiclassical.
		\subsection{Resolution of identity}
As far as the present work is concerned, the most important property that coherent states $\{|g, \vec{p}\rangle\}$ satisfy is that they form a (overcomplete) basis for the Hilbert space $\mathcal{H}$. This is neatly expressed by the resolution of identity operator, i.e.
\begin{align} \label{res-of-idy}
\mathbb{I} = \int_{\Gamma} d\mu(g, \vec{p}) | g, \vec{p} \rangle \langle g, \vec{p} |
\end{align}
for some measure $d\mu$. The measure which does the job is given explicitly by Eq. (4.82) in \cite{LQG-CS2}:
\begin{align}
d\mu(g, \vec{p}) = d\nu(\vec p) \ d\mu_H(g) = e^{-t/4}\biggl(\frac{t}{\pi}\biggr)^{3/2} \dfrac{\sinh tp}{tp} e^{-tp^2} \, d^3p \, d\mu_H(g)
\end{align}
where $p := |\vec{p}|$. The proof of (\ref{res-of-idy}) is based on Schur's lemma, a generalization of which is provided by Eq. (\ref{schur-general}) in appendix \ref{intertwiners}.
		\subsection{Behavior under gauge transformations}
For later use, let us establish how the coherent states $\ket{g_0,\vec p_0}$ transform under local gauge transformations of the holonomy. Under such a transformation, which is described by a function $a(x)\in SU(2)$, the holonomy itself transforms as
\be
g_l \to a_t g_l a_s^{-1},
\ee
where $a_t$ and $a_s$ denote $a(x)$ evaluated at the target and the source of the link $l$. Using this in the expression \eqref{CS-in-g} for the wave function of the coherent state, one finds that the state expressed in the left-representation transforms as
\be\label{gauge-left}
\ket{g_0,\vec p_0} \to \ket{a_t^{-1}g_0a_s, R(a_s^{-1})\vec p_0},
\ee
where $R(a)$ is the $\mathbb{R}^3$ rotation matrix associated with the element $a\in SU(2)$, defined by the relation $a(p^i\sigma_i)a^{-1} \equiv R(a)^i_{\;j} p^j$. For the state in the right-representation, one similarly has
\be\label{gauge-right}
\ket{g_0,\ppr_0} \to \ket{a_t^{-1}g_0a_s, R(a_t^{-1})\ppr_0}.
\ee
Thus we find that, while the coherent state $\ket{g_0,\vec p_0}$ is not gauge invariant, it transforms under a gauge transformation into a coherent state of the same form but peaked on a transformed phase space point.

		\subsection{Relation to classical variables}
The labels $g_0$ and $\vec p_0$, which parametrize the coherent state $\ket{g_0,\vec p_0}$, have a straightforward interpretation in terms of the classical variables on which loop quantum gravity is based. The group element $g_0$ is clearly to be identified with the holonomy of the Ashtekar--Barbero connection along the link $l$,
\be\label{c-holonomy}
h_l[A] = {\cal P}\exp\biggl(-\int_l A\biggr).
\ee
The conjugate variable is the flux of the densitized triad $E^a_i(x)$ through a surface dual to the link. To make the correspondence with $\vec p_0$ and $\ppr_0$ precise, one has to associate two such variables to the link, which is naturally done using the parallel transported flux variable,
\be\label{c-flux}
E^{[p]}(S) = \int_S d^2\sigma\,n_a(\sigma) h_{p\leftarrow\sigma}E^a(\sigma)h_{\sigma\leftarrow p}.
\ee
Here $n_a(\sigma)$ is a normal vector of the surface, $E^a(\sigma)$ is the $su(2)$-valued object $E^a(\sigma) = E^a_i(\sigma)\tau^i$, and $p$ denotes a point on the link $l$. The holonomy $h_{\sigma\leftarrow p}$ transports from the point $p$ to a point $\sigma$ along a path which follows the link up to the point where the link intersects the surface, and from there goes to $\sigma$ along the surface (e.g. along a straight line in the coordinates chosen on the surface). Choosing the point $p$ as the source or the target of the link, one obtains two flux variables, which are related by
\be
E^{[t(l)]} = h_lE^{[s(l)]}h_l^{-1}.
\ee
Comparison with Eq. \eqref{pandp'} now makes it clear that $\vec p$ and $\ppr$ correspond respectively to $E^{[s(l)]}$ and $E^{[t(l)]}$. Note also that under a gauge transformation, the variables \eqref{c-holonomy} and \eqref{c-flux} transform as
\begin{align}
h_l &\to a_{t(l)}h_la_{s(l)}^{-1} \\
E^{[s(l)]} &\to a_s E^{[s(l)]} a_s^{-1} \\
E^{[t(l)]} &\to a_t E^{[t(l)]} a_t^{-1}
\end{align}
which is consistent with the transformation of $g_0$, $\vec p_0$ and $\ppr_0$ given in Eqs. \eqref{gauge-left} and \eqref{gauge-right}.

		\subsection{Coherent states on a fixed graph}
Up to now we only considered coherent states on a single copy of $SU(2)$, i.e., corresponding to a single link in a graph. The generalization of coherent states to a graph $\Gamma$ with $L$ links is straightforward, at least at the gauge-variant level. Indeed, the Hilbert space associated to such a graph is simply
\begin{align} \label{HS-graph}
\mathcal{H} = \bigotimes_{l = 1}^{L} L_2(SU(2), d\mu_H) = L_2(SU(2)^L, d\mu_H)
\end{align}
where by $d\mu_H$ we really understand $\prod_{l = 1}^{L} d\mu_H$, with no risk of confusion if the argument of the measure is given. On this Hilbert space, one simply defines coherent states as the product of single-link coherent states: in particular, in the group and algebra representations we have
\begin{align}\label{product-cs}
\psi^{t}_{(\{g_0^l\}, \{\vec p_0^{\,l}\})}(g_1,\dots,g_L) \equiv \langle g_1,\dots,g_L | g_0^1,\vec p_0^1; \dots; g_0^L, \vec p_0^L\rangle = \prod_l \sum_{j_l} d_{j_l} e^{-t  \lambda_{j_l}/2} \chi^{(j_l)}(g_0^l e^{t\vec p_0^{\,l} \cdot \vec{\sigma}/2} g_l^{-1})
\end{align}
and
\begin{align} \label{cs-on-graph}
\ket{g_0^1,\vec p_0^1; \dots; g_0^L, \vec p_0^L} \equiv |\{g_0^l\}, \{\vec p_0^{\,l}\}\rangle = \prod_l \sum_{j_l, m_l, n_l} \sqrt{d_{j_l}} e^{-t\lambda_{j_l}/2} D^{(j_l)}_{m_l n_l}(g_0^l e^{t\vec p_0^{\,l} \cdot \vec{\sigma}/2}) |j_l, m_l, n_l\rangle
\end{align}
respectively. 

The states $|\{g_0^l\}, \{\vec p_0^{\,l}\}\rangle$ are not gauge-invariant. To obtain from them a family of gauge-invariant coherent states, one must consider the node structure of the graph $\Gamma$, and perform a group averaging at each of the $N$ nodes:
\begin{align} \label{gauge-invariance-def}
\Psi^{t}_{\Gamma,\{g_0^l\}, \{\vec p_0^{\,l}\}}(g_1,\dots,g_L) = \int_{SU(2)^N} d\mu_H(\tilde{g}_1) ... d\mu_H(\tilde{g}_N) \ \psi^{t}_{(\{g_0^l\}, \{\vec p_0^{\,l}\})}(\tilde{g}_{t(1)}^{-1} g_1 \tilde{g}_{s(1)}, \dots, \tilde{g}_{t(L)}^{-1} g_L \tilde{g}_{s(L)})
\end{align}
where we denote by $s(l)$ and $t(l)$ the source and the target nodes of the link $l$. The definition (\ref{gauge-invariance-def}) can be expanded to obtain
\begin{align}
\Psi^{t}_{\Gamma,\{g_0^l\}, \{\vec p_0^{\,l}\}}(g_1,\dots,g_L) &= \biggl(\prod_l \sum_{j_l} d_{j_l} e^{-t  \lambda_{j_l}/2} D^{(j_l)}_{m_l n_l}(g_0^l e^{t\vec p_0^{\,l} \cdot \vec{\sigma}/2})\overline{D^{(j_l)}_{\mu_l \nu_l}(g_l)}\biggr) \notag \\
&\times \int_{SU(2)^N} d\mu_H(\tilde{g}_1) ... d\mu_H(\tilde{g}_N)\, \biggl(\prod_l D^{(j_l)}_{\mu_l n_l}(\tilde g_{t(l)}) D^{(j_l)}_{n_l\nu_l}(\tilde g_{s(l)}^{-1})\biggr).\label{invpsi}
\end{align}
Here, as well as in the remainder of the article, summation over repeated indices is understood.

In Eq. \eqref{invpsi} the multiple group integral is effectively a projector onto the intertwiner space of the graph, and can be expressed in terms of intertwiners compatible with the structure of the graph; schematically, it has the form $\sum_\iota \ket\iota\bra\iota$. As the result, we find that the gauge invariant coherent state can be written in the form
\begin{align}
&\Psi^{t}_{\Gamma,\{g_0^l\}, \{\vec p_0^{\,l}\}}(g_1,\dots,g_L) \notag \\
&= \sum_{j_l,\iota_n} e^{-t(\lambda_{j_1} + \dots + \lambda_{j_L})/2}\,\overline{\Phi_{\Gamma,\{ j_l\},\{\iota_n\}}(g_0^1e^{t\vec p_0^{\,1}\cdot\vec\sigma/2},\dots,g_0^Le^{t\vec p_0^{\,L}\cdot\vec\sigma/2})}\,\Phi_{\Gamma,\{ j_l\},\{\iota_n\}}(g_1,\dots,g_L), \label{inv-cs}
\end{align}
where we have introduced the notation
\be\label{spinnetwork}
\Phi_{\Gamma,\{ j_l\},\{\iota_n\}}(g_1,\dots,g_L) = \biggl(\prod_n \iota_n\biggr)^{n_1\cdots n_L}_{m_1\cdots m_L} \biggl( \prod_l \sqrt{d_{j_l}}\overline{D^{(j_l)}_{m_ln_l}(g_l)}\biggr)
\ee
for a standard spin network state defined on the graph.


%
%
%
\section{Coherent state operators: General discussion}\label{CS-general}
\subsection{Construction of the operators}
We already spoke about coherent state operators, i.e., operators ``built'' from a given family of coherent states. It is now time that we show how this is done. The idea is based on the resolution of identity (\ref{res-of-idy}), that any family of coherent states must satisfy. Given any function $f(g,\vec p)$ on the classical phase space associated with a single link, one can construct an operator by inserting $f(g,\vec p)$ inside the integral in Eq. \eqref{res-of-idy}, thus obtaining the operator 
\be
f \ \ \longrightarrow \ \ \hat{A}_f := \int d\mu(g,p)\,f(g,\vec p)\,\ket{g,\vec p}\bra{g,\vec p}.
\ee
This procedure uniquely associates an operator on the single-link Hilbert space (\ref{HS-link}) to any phase space function $f$.

It is straightforward to generalize the construction to obtain operators acting on the Hilbert space \eqref{HS-graph} belonging to a given graph. To any function $f\bigl(\{g_l\},\{\vec p_l\}\bigr)$ on the phase space $\Gamma \approx SU(2)^L \times \mathbb{R}^{3L}$, one associates the operator
\begin{align} \label{cs-quantization-map}
f \ \ \longrightarrow \ \ \hat{A}_f = \int_\Gamma d\mu(g_1,\vec p_1)\cdots d\mu(g_L,\vec p_L)\, f\bigl(\{g_l\},\{\vec p_l\}\bigr)\, | \{g_l\},\{\vec p_l\} \rangle \langle \{g_l\},\{\vec p_l\}|.
\end{align}
Here $\bigl\{\ket{\{g_l\},\{\vec p_l\}}\bigr\}$ can be any set of coherent states that resolve the identity on the Hilbert space of the graph. In this work we will focus on operators that are obtained by choosing the state $\ket{\{g_l\},\{\vec p_l\}}$ as the simple (gauge-variant) tensor product of single-link coherent states given in Eqs. \eqref{product-cs} and \eqref{cs-on-graph}. On the one hand, the primary reason for making this choice is that it is the simplest one available; however, on the other hand one does not seem to gain much from carrying out the construction of operators using gauge invariant states instead. Specifically, one might fear that the following problems could arise as a result of our choice of non-gauge invariant states:
\begin{itemize}
\item It might be difficult to obtain gauge invariant operators, if one is constructing them using states which are not gauge invariant.
\item By using two different families (that is, gauge invariant vs. non-gauge invariant coherent states), one might associate two different operators to the same classical function.
\end{itemize}
The first concern is answered in the next section, where we present a general analysis of the gauge invariance of coherent state operators. It turns out that the prescription \eqref{cs-quantization-map} results in a gauge invariant operator whenever the function $f(\{g_l\}, \{\vec{p_l}\})$ is invariant under the corresponding transformation of its arguments. As for the second concern, while one should not necessarily expect different families of coherent states to produce the same operator, we actually find that repeating the construction with gauge invariant coherent states in Eq. \eqref{cs-quantization-map} does not actually change anything: the action of the resulting operator on gauge invariant states will be identical to that of the operator obtained from non-gauge invariant states. The proof of this statement is given in Appendix \ref{invstates}.

At this point we need to make a technical remark concerning the relation between the orientation of the graph and the variables $\vec p$ and $\ppr$. The classical interpretation of these variables indicates that $\vec p$ belongs to the source of its associated link, and $\ppr$ belongs to the target. This distinction is respected by the construction \eqref{cs-quantization-map}, in the sense that whenever one writes a $\vec p$ in the classical function $f\bigl(\{g_l\},\{\vec p_l\}\bigr)$, the resulting operator will act at the source of the corresponding link, whereas by writing a $\ppr$ for the same link, the resulting operator will act at the target of the link. This needs to be taken into account whenever one wants an operator which acts on the nodes of a spin network (such as the volume operator); in order to obtain such an operator, one has to write a $\vec p$ in the function $f$ for each link going out of the node, and a $\ppr$ for each link going into the node.

		\subsection{Gauge invariance of coherent state operators}
With the help of Eqs. \eqref{gauge-left} and \eqref{gauge-right}, it is straightforward to discuss the behaviour of the operator $\hat A_f$ of Eq. \eqref{cs-quantization-map} under gauge transformations. In the general case, where each link may be associated either with a $\vec p$ or with a $\ppr$, it is convenient to separate the two kinds of momentum variables, writing the operator as 
\be
\hat{A}_f = \int d\mu(g_1,\vec p_1)\cdots d\mu(g_L,\vec p_L)\, f\bigl(\{g_l\},\{\vec p_l\},\{\ppr_l\}\bigr)\, | \{g_l\},\{\vec p_l\},\{\ppr_l\} \rangle \langle \{g_l\},\{\vec p_l\},\{\ppr_l\}|,
\ee
with the understanding that in the measure each $\vec p_l$ stands for either $\vec p_l$ or $\ppr_l$. If we denote by $\hat U(a)$ the operator of gauge transformations, it follows from Eqs. \eqref{gauge-left} and \eqref{gauge-right} that
\be
\hat U(a)\ket{\{g_l\},\{\vec p_l\},\{\ppr_l\}} = \ket{\{a_{t(l)}^{-1}g_la_{s(l)}\},\{R(a_{s(l)}^{-1})\vec p_l\},\{R(a_{t(l)}^{-1})\ppr_l\}}.
\ee
Therefore we have
\begin{align}
\hat U(a)\hat A_f\hat U(a)^{-1} &= \int d\mu(g_1,\vec p_1)\cdots d\mu(g_L,\vec p_L) \notag \\
&{}\times f\bigl(\{a_{t(l)}g_la_{s(l)}^{-1}\},\{R(a_{s(l)})\vec p_l\},\{R(a_{t(l)})\ppr_l\}\bigr)\, | \{g_l\},\{\vec p_l\},\{\ppr_l\} \rangle \langle \{g_l\},\{\vec p_l\},\{\ppr_l\}|, \label{transformed-A}
\end{align}
where we have used the left and right invariance of the Haar measure $d\mu_H$, and the rotational invariance of the measure $d\nu$ to move the effect of the gauge transformations from the states to the function $f$.

From Eq. \eqref{transformed-A} it is clear that it is possible for the operator $\hat A_f$ to be gauge invariant, and the condition for its gauge invariance is that the function $f$ is invariant under the corresponding transformation induced on its arguments. In particular, if the function $f$ depends only on the $\vec p$ -variables associated to a single node $n$ -- which is the case for operators such as the angle operator and the volume operator -- the condition for gauge invariance reads
\be
f\bigl(\{R(a_n)\vec p_l\}\bigr) = f\bigl(\{\vec p_l\}\bigr),
\ee
i.e., that the function $f$ is invariant under a common rotation of all its arguments.
 	 \subsection{Operators corresponding to positive phase space functions}\label{positivity}
Another general property of the operator \eqref{cs-quantization-map}, which we will now establish, is the following: Suppose the function $f$ is strictly positive, $f\bigl(\{g_l\},\{\vec p_l\}\bigr)>0$, everywhere except possibly in a set of measure zero with respect to the measure $d\mu(g_1,\vec p_1)\cdots d\mu(g_L,\vec p_L)$. Then all eigenvalues of $\hat A_f$ (corresponding to proper, i.e. non-distributional eigenstates) are strictly positive, and the eigenvalue zero will not appear in the spectrum of the operator.

To prove this statement, suppose that $\ket\lambda$ is a (proper, normalized) eigenstate of $\hat A_f$ with eigenvalue $\lambda$. Then the eigenvalue can be expressed as
\be\label{lambda}
\lambda = \bra{\lambda}\hat A_f\ket{\lambda} = \int d\mu(g_1,\vec p_1)\cdots d\mu(g_L,\vec p_L)\, f\bigl(\{g_l\},\{\vec p_l\}\bigr)\,\bigl|\braket{\{g_l\},\{\vec p_l\}}{\lambda}\bigr|^2.
\ee
This shows that $\lambda$ is manifestly non-negative. If the set where $f\bigl(\{g_l\},\{\vec p_l\}\bigr)\leq 0$ is of measure zero, then $\lambda$ can be equal to zero only if the set where $\braket{\{g_l\},\{\vec p_l\}}{\lambda}\neq 0$ is also of measure zero. However, if $\ket\lambda$ is a proper, non-distributional state, its projections on the basis states $\ket{\{g_l\},\{\vec p_l\}}$ have finite values. This, together with the condition
\be
\int d\mu(g_1,\vec p_1)\cdots d\mu(g_L,\vec p_L)\, \bigl|\braket{\{g_l\},\{\vec p_l\}}{\lambda}\bigr|^2 = 1,
\ee
is enough to ensure that the set where $\braket{\{g_l\},\{\vec p_l\}}{\lambda}\neq 0$ has positive measure. This implies that the integral on the right-hand side of Eq. \eqref{lambda} is strictly positive.
\section{Coherent state operators: Examples} \label{CS-examples}
\subsection{Holonomy operator}

The coherent state operator corresponding to the holonomy is constructed by taking the Wigner matrix $D^{(j)}_{mn}(g)$ as the function $f(g,p)$. In this way we obtain the operator
\be\label{A_D}
\hat A_{D^{(j)}_{mn}} = \int d\mu(g,p)\,D^{(j)}_{mn}(g)\,\ket{g,\vec p}\bra{g,\vec p},
\ee
which acts on the Hilbert space \eqref{HS-link} of a single link. In order to understand the action of this operator, we compute its matrix elements between two states of the basis $\{\ket{j,m,n}\}$. To begin, we recall the expression \eqref{CS-in-sn} for the coherent states, and obtain
\begin{align}
\bra{j_1,m_1,n_1} \hat A_{D^{(j)}_{mn}} \ket{j_2,m_2,n_2} &= \sqrt{d_{j_1}d_{j_2}}e^{-t(\lambda_{j_1}+\lambda_{j_2})/2}\int d\mu_H(g)\,D^{(j_1)}_{m_1\mu_1}(g)D^{(j)}_{mn}(g)\overline{D^{(j_2)}_{m_2\mu_2}(g)}\notag \\
&{}\times\int d\nu(\vec p)\,D^{(j_1)}_{\mu_1n_1}(e^{t\vec p\cdot\vec\sigma/2})\overline{D^{(j_2)}_{\mu_2n_2}(e^{t\vec p\cdot\vec\sigma/2})},
\end{align}
Here the integral over the group gives $(1/d_{j_2})C^{j_1jj_2}_{m_1mm_2}C^{j_1jj_2}_{n_1nn_2}$, while the integral over $\vec p$ can be evaluated by  coupling the two $D$-matrices by means of the relation \eqref{newCGseries}, and then using Eq. \eqref{schur-1} to compute the integral
\begin{align}
\int d\nu(\vec p)\,D^{(k)}_{mn}(e^{t\vec p\cdot\vec\sigma/2}) &= \delta_{mn}\frac{1}{d_k}\int d\nu(\vec p)\,\chi^{(k)}(e^{t\vec p\cdot\vec\sigma/2}) \notag \\
&=\delta_{mn}\frac{1}{d_k}e^{t(\lambda_k/4-1/8)}\biggl(d_k\cosh\frac{d_kt}{8} + \sinh\frac{d_kt}{8}\biggr).
\end{align}
In this way we get
\be
\bra{j_1,m_1,n_1} \hat A_{D^{(j)}_{mn}} \ket{j_2,m_2,n_2} = C^{j_1jj_2}_{m_1m m_2}\sum_k \frac{1}{d_k}B_t(k)e^{t(\lambda_k-2\lambda_{j_1}-2\lambda_{j_2})/4}C^{j_1j_2k}_{n_1\nu_1\mu_2}C^{j_1j_2k}_{\mu_1n_2\mu_2}C^{j_1jj_2}_{\mu_1n\nu_1},
\ee
where 
\be
B_t(k) = e^{-t/8}\biggl(d_k\cosh\frac{d_kt}{8} + \sinh\frac{d_kt}{8}\biggr).
\ee
To complete the calculation, we need to evaluate the contraction of three Clebsch-Gordan coefficients, which is conveniently done using graphical techniques; see Appendix \ref{intertwiners} and in particular Eqs. \eqref{cg-picture}, \eqref{3block} and \eqref{6j-picture}. As the result, we find
\begin{align}
\bra{j_1,m_1,n_1} \hat A_{D^{(j)}_{mn}} \ket{j_2,m_2,n_2} &= \sqrt{\frac{d_{j_1}}{d_{j_2}}}C^{j_1jj_2}_{m_1m m_2}C^{j_1jj_2}_{n_1nn_2}\sum_k (-1)^{2k}B_t(k)e^{t(\lambda_k-2\lambda_{j_1}-2\lambda_{j_2})/4}\begin{Bmatrix} j_1&j_2&j \\ j_1&j_2&k \end{Bmatrix} \notag \\
&= H_t(j_1,j_2,j)\bra{j_1,m_1,n_1} \hat D^{(j)}_{mn} \ket{j_2,m_2,n_2},
\end{align}
where we have recognized the matrix element of the canonical holonomy operator, given in Eq. \eqref{standard-holo-me}, and denoted the multiplicative factor as
\be
H_t(j_1,j_2,j) = \sum_{k} (-1)^{2k}e^{t(\lambda_{k}/4 - \lambda_{j_1}/2 - \lambda_{j_2}/2 - 1/8)}\left(d_{k}\cosh\frac{d_{k} t}{8}+ \sinh\frac{d_{k} t}{8}\right)\begin{Bmatrix} j_1&j_2&j \\ j_1&j_2&k \end{Bmatrix}.
\ee
The matrix elements of coherent state holonomy operator differ from those of the canonical operator by this factor. However, in the limit $t\to 0$ the factor $H_t(j_1,j_2,j)$ reduces to
\be
H_{t=0}(j_1,j_2,j) = \sum_k d_k(-1)^{2k}\begin{Bmatrix} j_1&j_2&j \\ j_1&j_2&k \end{Bmatrix} = 1,
\ee
where we evaluated the sum by setting $l=0$ in Eq. \eqref{6jsum} and using the explicit expression \eqref{6jzero} for a 6$j$-symbol with one argument equal to zero. Hence the canonical holonomy operator is recovered from the coherent state operator in the limit where the coherent states become sharply peaked in the group element.

\subsection{Left- and right-invariant vector fields}\label{vectorfields}

A natural candidate for the function $f$ which gives rise to the left-invariant vector field operator is the variable $\vec p$, which arises from the decomposition $h = ge^{t\vec p\cdot\vec\sigma/2}$, and which evidently is invariant under left multiplication by $SU(2)$. Indeed, the coherent state operator corresponding to the left-invariant vector field is
\be\label{A_p}
\hat A_{p^i} = -i\int d\mu(g,p)\,p^i\ket{g,\vec p}\bra{g,\vec p}.
\ee
We again study the action of this operator by computing its matrix elements in the spin network basis. To start the calculation, we need to express $p^i$ in terms of objects compatible with recoupling theory. This is achieved by writing
\be
p^i = \frac{|\vec p|}{2\sinh t|\vec p|}\,{\rm Tr}\,\bigl(\sigma^i e^{t\vec p\cdot\vec\sigma}\bigr).
\ee
Then, using again Eq. \eqref{CS-in-sn} for the coherent states in the spin network basis, we obtain
\begin{align}
\bra{j_1,m_1,n_1} \hat A_{p^i} \ket{j_2,m_2,n_2} &= -\frac{i}{2}\sqrt{d_{j_1}d_{j_2}}e^{-t(\lambda_{j_1}+\lambda_{j_2})}\sigma^i_{AB}\int d\mu_H(g)\,D^{(j_1)}_{m_1\mu_1}(g)\overline{D^{(j_2)}_{m_2\mu_2}(g)} \notag \\
&{}\times\int d\nu(\vec p)\,\frac{|\vec p|}{2\sinh t|\vec p|}D^{(1/2)}_{BA}(e^{t\vec p\cdot\vec\sigma})D^{(j_1)}_{\mu_1n_1}(e^{t\vec p\cdot\vec\sigma/2})\overline{D^{(j_2)}_{\mu_2 n_2}(e^{t\vec p\cdot\vec\sigma/2})},
\end{align}
where the integral over the group immediately gives $(1/d_{j_1})\delta_{j_1j_2}\delta_{m_1m_2}\delta_{\mu_1\mu_2}$. In the integral over $\vec p$, we then use Eq. \eqref{newCGseries} to couple the matrices $D^{(1/2)}_{BA}(e^{t\vec p\cdot\vec\sigma})$ and $D^{(j_1)}_{n_2n_1}(e^{t\vec p\cdot\vec\sigma})$, reducing the integral to
\begin{align}
\int d\nu(\vec p)\,\frac{|\vec p|}{\sinh t|\vec p|}D^{(j)}_{\mu_2\mu_1}(e^{t\vec p\cdot\vec\sigma}) &= \delta_{\mu_1\mu_2}\frac{1}{d_j}\int d\nu(\vec p)\,\frac{|\vec p|}{\sinh t|\vec p|}\chi^{(j)}(e^{t\vec p\cdot\vec\sigma}) \notag \\
&= \delta_{\mu_1\mu_2}\frac{2}{td_j}e^{-t/4}\sum_{s=-j}^j \bigl(1+2s^2t\bigr)e^{s^2t}.
\end{align}
At this point we are left with
\be\label{A_p step}
\bra{j_1,m_1,n_1} \hat A_{p^i} \ket{j_2,m_2,n_2} = -\frac{i}{2t}e^{-t(\lambda_{j_1}+1/4)}\delta_{j_1j_2}\delta_{m_1m_2}\sum_j \frac{1}{d_j}\sum_{s=-j}^j \bigl(1+2s^2t\bigr)e^{s^2t}\sigma^i_{AB}C^{\half j_1j}_{An_1\mu}C^{\half j_1j}_{Bn_2\mu}.
\ee
The contraction of Clebsch--Gordan coefficients can be evaluated by a graphical calculation as
\be
\sigma^i_{AB}C^{\half j_1j}_{An_1\mu}C^{\half j_1j}_{Bn_2\mu} = \sqrt{\frac{3}{2}}\frac{d_j}{\sqrt{d_{j_1}\lambda_{j_1}}}(-1)^{j-j_1+1/2}\begin{Bmatrix} j_1&j_1&1 \\ \half&\half&j\end{Bmatrix} {}^{(j_1)}\sigma^i_{n_2n_1}.
\ee
Inserting this back into Eq. \eqref{A_p step}, we see that due to the triangular conditions of the 6$j$-symbol, the sum over $j$ reduces to two terms, $j=j_1-\half$ and $j=j_1+\half$. The sum can then be evaluated explicitly, using the expressions \eqref{6jhalf} for the relevant 6$j$-symbols. In the end we find
\be
\bra{j_1,m_1,n_1} \hat A_{p^i} \ket{j_2,m_2,n_2} = \frac{i}{2}\delta_{j_1j_2}\delta_{m_1m_2}F_t(j_1){}^{(j_1)}\sigma^i_{n_2n_1},
\ee
that is,
\be\label{CS-left}
\hat A_{p^i} \ket{j,m,n} = F_t(j)\hat L^i \ket{j,m,n},
\ee
where $\hat L^i$ is the standard left-invariant vector field, whose action is given in Eq. \eqref{standard-flux-me}, and the multiplicative factor is given by
\be\label{F_t}
F_t(j) = \dfrac{1}{2 t d_{j} \lambda_{j}} \biggl[j \left(d_{j}^2 t + 2\right) - e^{-d_{j}^2 t/4}\sum_{s = -j + \frac{1}{2}}^{j - \frac{1}{2}} \bigl(1+2s^2t\bigr) e^{s^2t}\biggr].
\ee
We may check that for large spins, the asymptotic behavior of the factor is $F_t(j) = 1 + {\cal O}(1/j)$, independently of the value of $t$. Hence the operator \eqref{CS-left} behaves approximately like the canonical left-invariant vector field, when it is applied on a state $\ket{j,m,n}$ with $j\gg 1$.

The coherent state operator of the right-invariant vector field is obtained similarly, but instead of $\vec p$ we use the variable $\ppr$, corresponding to the decomposition $h = e^{t\ppr\cdot\vec\sigma/2}g$. The operator is given by
\be
\hat A_{(p')^i} = i\int d\mu(g,p')\,(p')^i\ket{g,\ppr}\bra{g,\ppr},
\ee
and its action on spin network states is entirely similar to that of the left-invariant vector field:
\be\label{CS-right}
\hat A_{(p')^i}\ket{j,m,n} = F_t(j)\hat R^i\ket{j,m,n},
\ee
where $\hat R^i$ is the canonical operator from Eq. \eqref{standard-R-me}.

With the help of Eqs. \eqref{CS-left} and \eqref{CS-right}, we can now confirm the statement made earlier, that the variables $\vec p$ and $\ppr$ give rise to operators acting respectively on the source and the target of the corresponding link. Our convention for the indices of the holonomy is shown in Eq. \eqref{Dmatrix}; the indices $m$ and $n$ in $D^{(j)}_{mn}(g_l)$ are associated respectively with the target and the source of the link $l$. If we now write explicitly the action of the coherent state left- and right-invariant vector fields on the state $\ket{j,m,n}$,
\begin{align}
\hat A_{p^i}\ket{j,m,n} &= \frac{i}{2}F_t(j){}^{(j)}\sigma^i_{n'n}\ket{j,m,n'}, \\
\hat A_{(p')^i}\ket{j,m,n} &= -\frac{i}{2}F_t(j){}^{(j)}\sigma^i_{m'm}\ket{j,m',n},
\end{align}
we see that $\hat A_{p^i}$ indeed acts on the index belonging to the source, and $\hat A_{(p')^i}$ acts on the index belonging to the target.

\subsection{Algebra of holonomies and fluxes}

Having derived the action of the coherent state holonomy and flux operators in the basis $\{\ket{j,m,n}\}$, we may now study the algebra of these operators. By comparing the result with the commutation relation of the corresponding canonical operators,
\be
[\hat D^{(j)}_{mn},\hat L^i] = \frac{i}{2}{}^{(j)}\sigma^i_{\mu n}\hat D^{(j)}_{m\mu}, \\
\ee
we will see explicitly that the quantization of holonomies and fluxes by the coherent states prescription is not equivalent to their standard canonical quantization. 

Since we have the action of the coherent state operators at the level of matrix elements, we wish to compute $\bra{j_1,m_1,n_1}[\hat A_{D^{(j)}_{mn}},\hat A_{L^i}]\ket{j_2,m_2,n_2}$ and compare it with $\bra{j_1,m_1,n_1}\tfrac{i}{2}{}^{(j)}\sigma^i_{m'n}\hat A_{D^{(j)}_{mm'}}\ket{j_2,m_2,n_2}$. To evaluate the matrix element of the commutator, we begin by inserting a resolution of identity in the spin network basis. The resulting expression can be written as
\begin{align}
&\bra{j_1,m_1,n_1}[\hat A_{D^{(j)}_{mn}},\hat A_{L^i}]\ket{j_2,m_2,n_2} = \frac{i}{2}H_t(j_1,j_2,j)\sqrt{\frac{d_{j_1}}{d_{j_2}}}C^{j_1jj_2}_{m_1m m_2}\Bigl({}^{(j_2)}\sigma^i_{n_2\mu}C^{j_1jj_2}_{n_1n\mu} - C^{j_1jj_2}_{\mu nn_2}{}^{(j_1)}\sigma^i_{\mu n_1}\Bigr) \notag \\
&{}+ \frac{i}{2}H_t(j_1,j_2,j)\sqrt{\frac{d_{j_1}}{d_{j_2}}}C^{j_1jj_2}_{m_1m m_2}\Bigl(\bigl(F_t(j_2)-1\bigr){}^{(j_2)}\sigma^i_{n_2\mu}C^{j_1jj_2}_{n_1n\mu} - \bigl(F_t(j_1)-1\bigr)C^{j_1jj_2}_{\mu nn_2}{}^{(j_1)}\sigma^i_{\mu n_1}\Bigr).
\end{align}
To deal with the terms on the first line, we express the Clebsch--Gordan coefficients in terms of 3$j$-symbols, and use the relation \eqref{3j-invariance} to show that ${}^{(j_2)}\sigma^i_{n_2\mu}C^{j_1jj_2}_{n_1n\mu} - C^{j_1jj_2}_{\mu nn_2}{}^{(j_1)}\sigma^i_{\mu n_1} = C^{j_1jj_2}_{n_1\mu n_2}{}^{(j)}\sigma^i_{\mu n}$. Then, reintroducing the matrix elements of the operator $\hat A_{D^{(j)}_{mn}}$, we find
\begin{align}
\bra{j_1,m_1,n_1}[\hat A_{D^{(j)}_{mn}},\hat A_{L^i}]\ket{j_2,m_2,n_2} &= \bra{j_1,m_1,n_1}\frac{i}{2}{}^{(j)}\sigma^i_{\mu n}\hat A_{D^{(j)}_{m\mu }} \ket{j_2,m_2,n_2} \notag \\
&{}+\frac{i}{2}\bigl(F_t(j_2)-1\bigr){}^{(j_2)}\sigma^i_{n_2\mu }\bra{j_1,m_1,n_1}\hat A_{D^{(j)}_{mn}}\ket{j_2,m_2,\mu } \notag \\
&{}-\frac{i}{2}\bigl(F_t(j_1)-1\bigr){}^{(j_1)}\sigma^i_{\mu n_1}\bra{j_1,m_1,\mu }\hat A_{D^{(j)}_{mn}}\ket{j_2,m_2,n_2}. \label{CS-commutator}
\end{align}
The first term on the right is what one would expect based on the commutation relation of the canonical operators, but the presence of the additional terms means that the algebra of the coherent state operators is indeed not canonical. However, since the factor $F_t(j)$ approaches 1 for large $j$, the operators $\hat A_{D^{(j)}_{mn}}$ and $\hat A_{L^i}$ do approximately satisfy the canonical commutation relation, when they are applied to a state $\ket{j'm'n'}$ with $j'\gg j$.

\subsection{Area operator}

The area of a surface $S$ can be expressed in terms of the flux variable $E_i(S)$ associated with the surface as $\sqrt{E_i(S)E_i(S)}$, i.e. as the ''length'' of the flux variable. Accordingly, we define the coherent state operator of area associated to a link as
\be\label{CS-area}
\hat A_{|\vec p|} = \int d\mu(g,p)\,|\vec p|\ket{g,\vec p}\bra{g,\vec p},
\ee
The computation of the matrix elements of this operator between two spin network states is straightforward. Proceeding as we did with the operator $\hat A_{p^i}$ in section \ref{vectorfields}, we obtain
\be\label{area-mat.el.}
\bra{j_1,m_1,n_1} \hat A_{|\vec p|} \ket{j_2,m_2,n_2} = e^{-t\lambda_{j_1}}\delta_{j_1j_2}\delta_{m_1m_2}\int d\nu(\vec p)\,|\vec p|D^{(j_1)}_{n_2n_1}(e^{t\vec p\cdot\vec\sigma}),
\ee
where the integral over $\vec p$ simply gives a factor proportional to $\delta_{n_1n_2}$. Hence we conclude that the operator \eqref{CS-area} is diagonal on the states $\ket{j,m,n}$,
\be
\hat A_{|\vec p|} \ket{j,m,n} = \alpha(j)\ket{j,m,n},
\ee
and evaluation of the integral in Eq. \eqref{area-mat.el.} shows that the eigenvalue is given by
\be\label{areaeig}
\alpha(j) = \left(\frac{1}{d_jt} + \frac{d_j}{2}\right){\rm erf}\left(\frac{\sqrt td_j}{2}\right) + \frac{1}{\sqrt{\pi t}}e^{-(\lambda_j+1/4)t},
\ee
where the error function is defined as ${\rm erf}(x) = \tfrac{2}{\sqrt\pi}\int_0^x dt\,e^{-t^2}$.

The eigenvalues \eqref{areaeig} are shown in Fig. \ref{areaplot} for various values of $t$. We see that
\begin{itemize}
\item For large $j$ the eigenvalues of the coherent state operator converge to the canonical eigenvalues. The asymptotic behaviour of the coherent state eigenvalue is $\alpha(j) = j + {\cal O}(1)$, independently of the value of $t$. The convergence is faster for larger values of $t$, reflecting the fact that as the value of $t$ is increased, the coherent states become more sharply peaked on the momentum variable.
\item For sufficiently small spins $\alpha(j)$ deviates significantly from the corresponding canonical eigenvalue. The difference is the most significant for $j=0$, as the lowest eigenvalue $\alpha(0)$ of the coherent state operator is always positive, which could have been anticipated on grounds of the theorem of section \ref{positivity} -- in fact $\alpha(0) > \half$ for any value of $t$. Hence even a link of spin zero carries a non-zero area according to the operator \eqref{CS-area}.
\end{itemize}

%
%

\subsection{Angle operator}

Consider two links belonging to a node of a spin network. An operator describing the angle between the corresponding flux vectors can be defined as
\be\label{CS-angle}
\hat A_{\theta(\vec p_1,\vec p_2)} = \int d\mu(g_1,p_1)\,d\mu(g_2,p_2)\,\cos^{-1}\biggl(\frac{\vec p_1\cdot\vec p_2}{|\vec p_1||\vec p_2|}\biggr)\,|g_1,\vec p_1;g_2,\vec p_2\rangle\langle g_1,\vec p_1;g_2,\vec p_2|.
\ee
Here we have assumed that both links are going out of the node. According to the discussion in section \ref{CS-general}, we should replace $\vec p$ with $\ppr$ for every link coming in to the node.

Let us compute the action of this operator on a spin network state where the two links have spins $j_1$ and $j_2$, and are coupled to a total spin $k$. The relevant (normalized) part of the state has the form
\be\label{j1j2k-state}
\ket{\Psi^{(j_1j_2;k)}_{m_1m_2;\alpha}} = \Biggl|\makeSymbol{\includegraphics[scale=1.25]{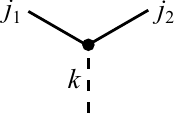}}\Biggr\rangle \equiv \sqrt{d_k} (-1)^{j_1-n_1+j_2-n_2}\begin{pmatrix} j_1&j_2&k \\ -n_1&-n_2&\alpha \end{pmatrix}|j_1m_1n_1\rangle|j_2m_2n_2\rangle.
\ee
As before, we find the action of the operator by computing its matrix elements between two states of this form. Using once again Eq. \eqref{CS-in-sn} for the coherent states, and carrying out the immediate group integrations, we are left with
\begin{align}
\bra{\Psi^{(j_1j_2;k)}_{m_1m_2;\alpha}}\hat A_\theta \ket{\Psi^{(j_1'j_2';l)}_{n_1n_2;\beta}} &= \delta_{j_1j_1'}\delta_{j_2j_2'}\delta_{m_1n_1}\delta_{m_2n_2}e^{-t(\lambda_{j_1}+\lambda_{j_2})} \notag \\
&{}\times \sqrt{d_kd_l}(-1)^{j_1-\mu_1+j_2-\mu_2}(-1)^{j_1-\nu_1+j_2-\nu_2}\begin{pmatrix} j_1&j_2&k \\ -\mu_1&-\mu_2&\alpha \end{pmatrix} \begin{pmatrix} j_1&j_2&l \\ -\nu_1&-\nu_2&\beta \end{pmatrix} \notag \\
&{}\times \int d\nu(\vec p_1)\,d\nu(\vec p_2)\,\cos^{-1}\biggl(\frac{\vec p_1\cdot\vec p_2}{|\vec p_1||\vec p_2|}\biggr) D^{(j_1)}_{\nu_1\mu_1}(e^{t\vec p_1\cdot\vec\sigma})D^{(j_2)}_{\nu_2\mu_2}(e^{t\vec p_2\cdot\vec\sigma}).\label{angle-step}
\end{align}
Here the integral on the last line, denote it by $I_{\nu_1\nu_2;\mu_1\mu_2}(j_1,j_2)$, is of the type discussed in Appendix \ref{expanding} -- see Eq. \eqref{schur-general} -- and can be written as
\be
I_{\nu_1\nu_2;\mu_1\mu_2}(j_1,j_2) = \sum_x d_x\,I(j_1,j_2,x)\iota^{(x)}_{\nu_1\nu_2;\mu_1\mu_2},
\ee
where
\be
\iota^{(x)}_{\nu_1\nu_2;\mu_1\mu_2} = (-1)^{j_1-\nu_1+j_2-\nu_2}\begin{pmatrix} j_1&j_2&x \\ -\nu_1&-\nu_2&m \end{pmatrix} (-1)^{x-m} \begin{pmatrix} x&j_1&j_2 \\ -m&\mu_1&\mu_2 \end{pmatrix}
\ee
and
\be
I(j_1,j_2,x) = I_{\nu_1\nu_2;\mu_1\mu_2}(j_1,j_2)\iota^{(x)}_{\mu_1\mu_2;\nu_1\nu_2} = I_{\nu_1\nu_2;\mu_1\mu_2}(j_1,j_2)\begin{pmatrix} j_1&j_2&x \\ \nu_1&\nu_2& m \end{pmatrix} \begin{pmatrix} j_1&j_2&x \\ \mu_1&\mu_2& m\end{pmatrix}.
\ee
Inserting this back into Eq. \eqref{angle-step}, and using the orthogonality relation of the 3$j$-symbols, we deduce that the state \eqref{j1j2k-state} is an eigenstate of the operator \eqref{CS-angle},
\be
\hat A_{\theta(\vec p_1,\vec p_2)}\Biggl|\makeSymbol{\includegraphics[scale=1.25]{j1j2k-node.pdf}}\Biggr\rangle = \theta(j_1,j_2,k)\Biggl|\makeSymbol{\includegraphics[scale=1.25]{j1j2k-node.pdf}}\Biggr\rangle,
\ee
and the eigenvalue is given by
\begin{align}\label{CS-theta}
\theta(j_1,j_2,k) & = e^{-t(\lambda_{j_1}+\lambda_{j_2})}\int d\nu(p_1)\,d\nu(p_2)\,\cos^{-1}\left(\frac{\vec p_1\cdot\vec p_2}{|\vec p_1||\vec p_2|}\right) \notag \\ & \times \begin{pmatrix} j_1\;\;j_2\;\;k \\ m_1\,m_2\,\mu \end{pmatrix} D^{(j_1)}_{m_1n_1}(e^{\vec p_1\cdot\vec\sigma})D^{(j_2)}_{m_2n_2}(e^{\vec p_2\cdot\vec\sigma})\begin{pmatrix} j_1\;j_2\;k \\ n_1\,n_2\,\mu\end{pmatrix}.
\end{align}

By numerically evaluating the integral \eqref{CS-theta}, the coherent state angle operator may be compared with the canonical angle operator, whose eigenvalue $\theta_{\rm can}(j_1,j_2,k)$ on the state \eqref{j1j2k-state} is
\be\label{can-theta}
\cos\theta_{\rm can}(j_1,j_2,k) = \frac{k(k+1) - j_1(j_1+1) - j_2(j_2+1)}{2\sqrt{j_1(j_1+1)}\sqrt{j_2(j_2+1)}}.
\ee
In Figures \ref{angleplot} and \ref{angleplot-2}, we show the results of a numerical calculation of the eigenvalues of the coherent state operator in two different cases.

In Fig. \ref{angleplot} we have the ``equilateral'' case, in which the spins $j_1$, $j_2$ and $k$ are all equal to a common value $j$. For each value of $j$, we compute the eigenvalue of the coherent state angle operator for various values of the parameter $t$. We find that the eigenvalues seem to converge to certain values as the value of $t$ increases, and that the limiting value is reached the more rapidly, the larger the value of $j$ is. (The fluctuations in the eigenvalues for given $j$ and varying $t$, which are seen in the plot for large spins, can be attributed to numerical error, instead of being a genuine feature of the data.) The canonical eigenvalue for the equilateral case is independent of $j$, and is equal to $\theta_{\rm can}(j,j,j) = 2\pi/3$; we see that with increasing $j$, the eigenvalues of the coherent state operator approach the canonical eigenvalue.

In Fig. \ref{angleplot-2} we show the ``degenerate'' case, in which $j_1=j_2\equiv j$ and $k=2j$. The value of $t$ has been fixed to $t=3$. The canonical eigenvalue is now given by $\theta_{\rm can}(j,j,2j) = \cos^{-1}(j/(j+1))$. The general behaviour of the eigenvalues of the coherent state operator as a function of $j$ is similar to that of the canonical eigenvalues, even though the coherent state eigenvalues now approach the canonical eigenvalues with increasing $j$ much more slowly than in the equilateral case. However, the relative difference between the two sets of eigenvalues remains roughly constant as $j$ increases, as shown in Fig. \ref{angle-diff}. Therefore it seems that the eigenvalues of the coherent state operator approach zero in the limit of large spins, thus agreeing with the canonical eigenvalue in this limit.

\afterpage{

\begin{figure}[t]
	\centering
		\includegraphics[width=0.65\textwidth]{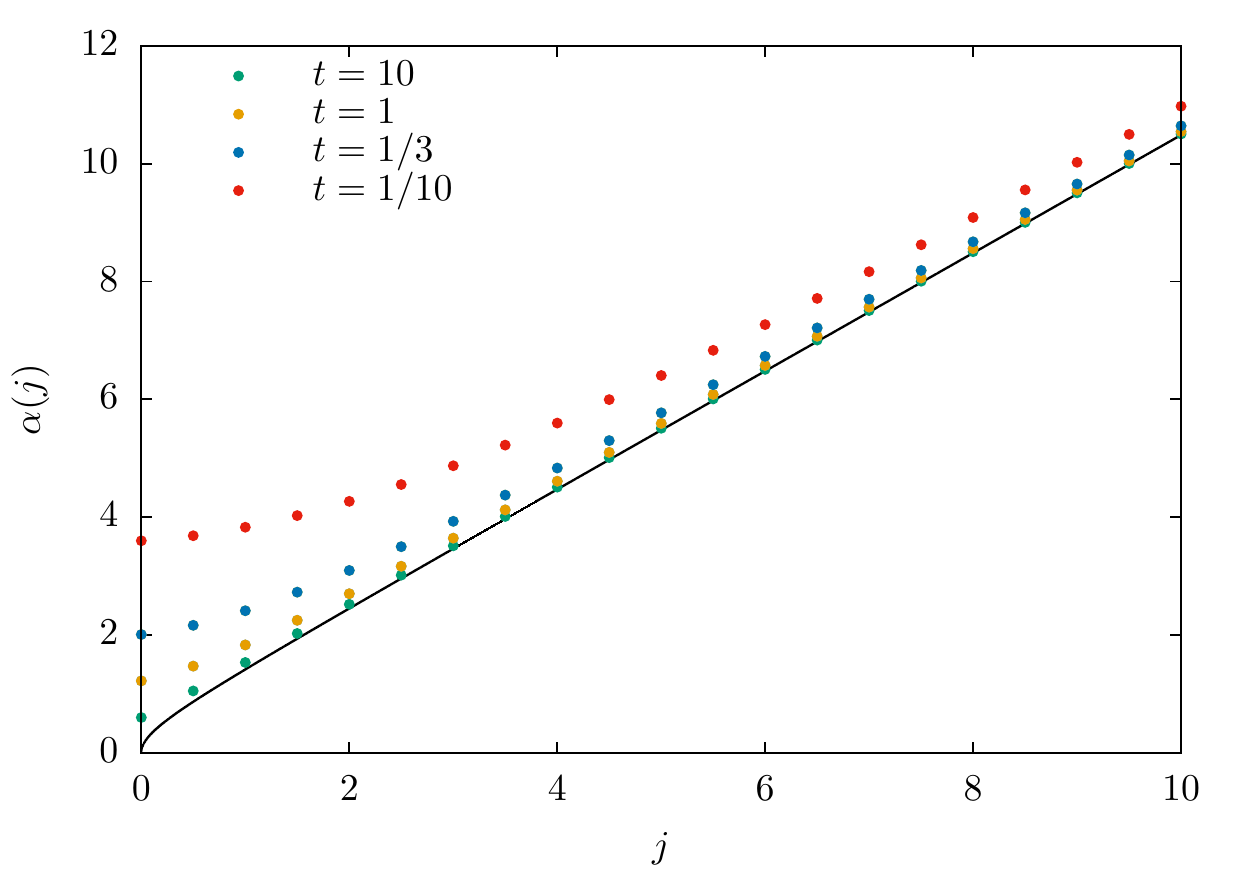}
		\caption{Eigenvalues of the coherent state area operator for some values of $t$. For comparison, the solid line shows the eigenvalues of the canonical area operator $A_{\rm can}(j) = \sqrt{j(j+1)}$. For large spins, the eigenvalues of the coherent state operator converge to the canonical eigenvalues.}
	\label{areaplot}
\end{figure}

\begin{figure}[b]
	\centering
		\includegraphics[width=0.65\textwidth]{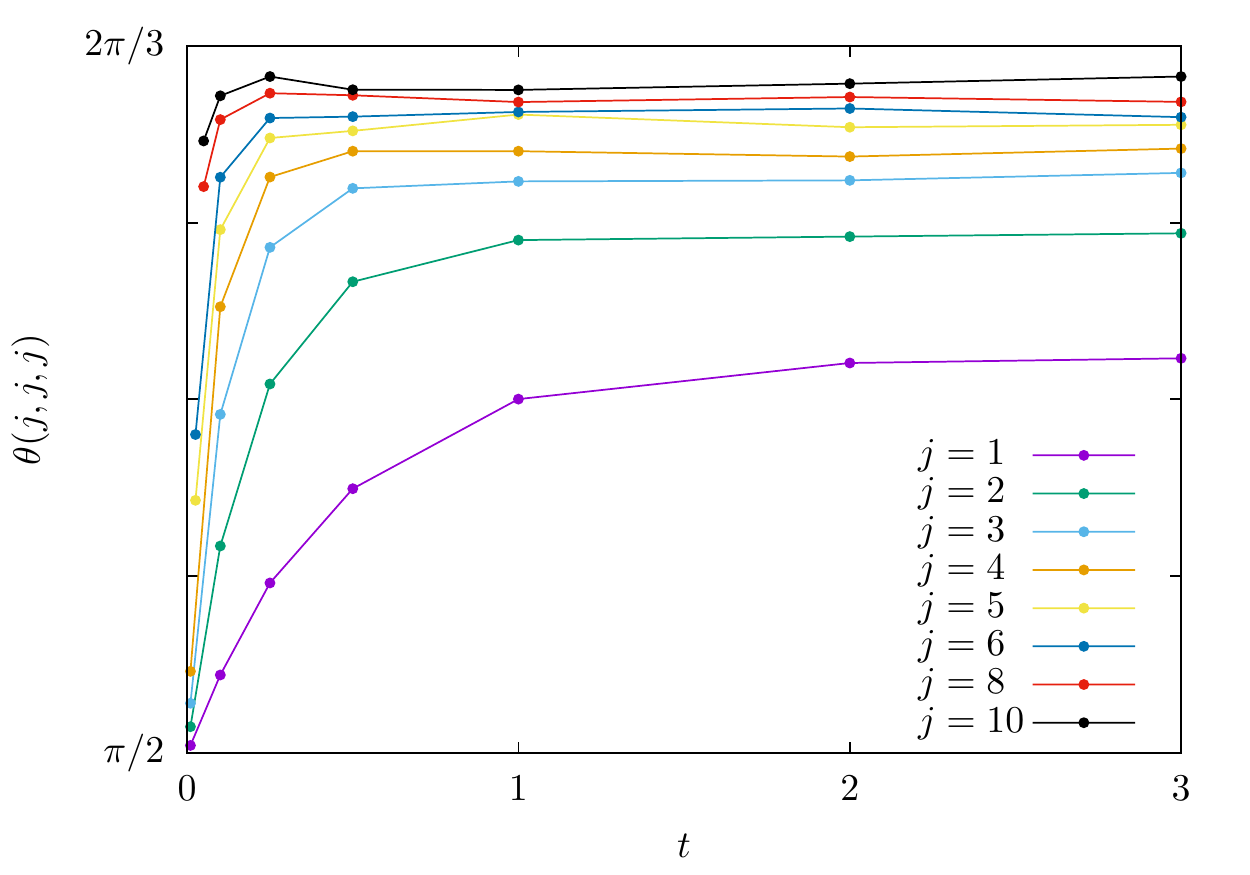}
		\caption{Numerical evaluation of eigenvalues of the coherent state angle operator in the ''equilateral'' case ($j_1=j_2=k\equiv j$) for various values of the parameter $t$. The corresponding eigenvalue of the canonical angle operator is equal to $2\pi/3$ independently of the value of $j$. As $j$ increases, the eigenvalues of the coherent state operator approach the canonical eigenvalue. The apparent fluctuations in the eigenvalues for large spins are due to numerical inaccuracy.}
	\label{angleplot}
\end{figure}

\newpage

\begin{figure}[t]
	\centering
		\includegraphics[width=0.65\textwidth]{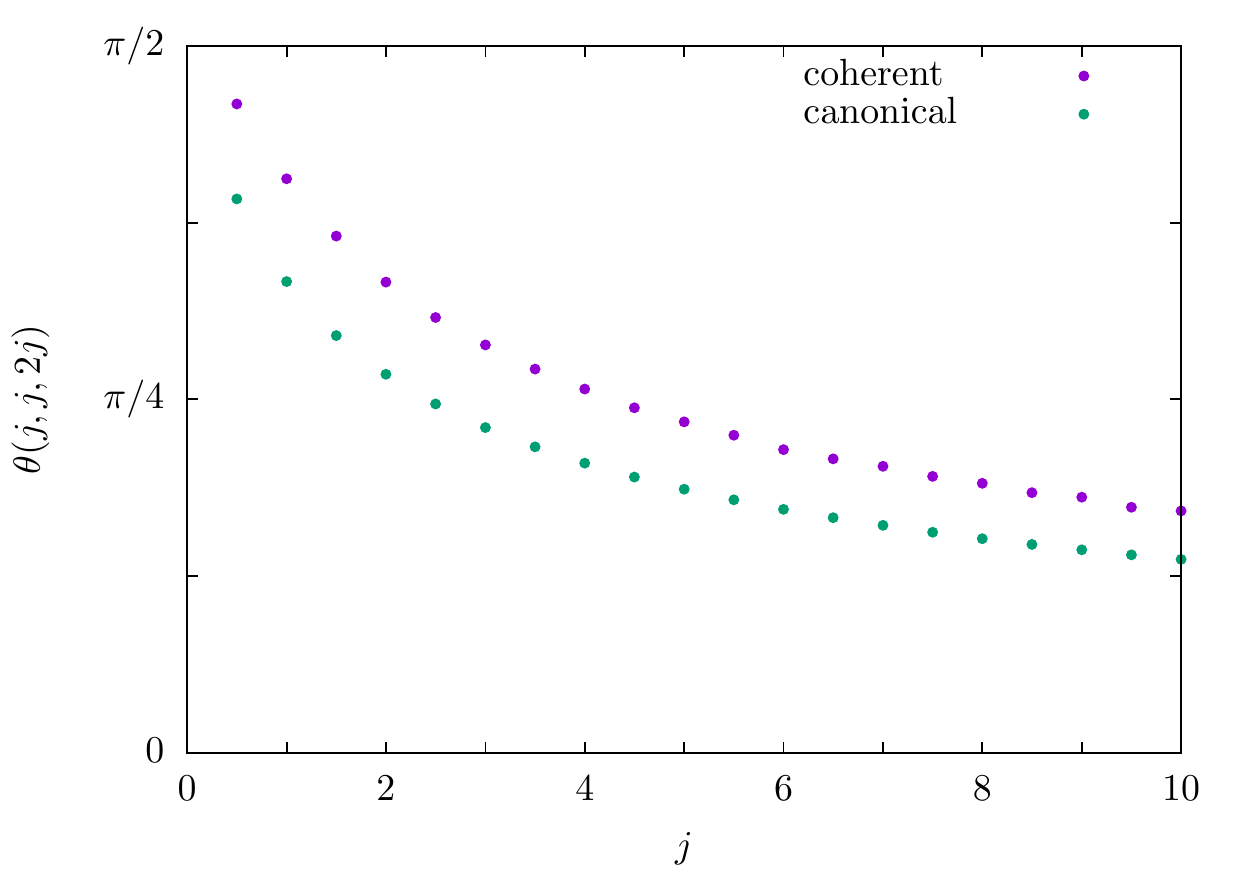}
		\caption{Numerical evaluation of eigenvalues of the coherent state angle operator in the ''degenerate'' case ($j_1=j_2\equiv j$ and $k=2j$) for $t=3$, and comparison with the corresponding canonical eigenvalues. The coherent state eigenvalues approach the canonical eigenvalues with increasing $j$ much more slowly than in the equilateral case. Even so, the relative difference between the two sets of eigenvalues is approximately constant as $j$ increases (see Fig. \ref{angle-diff}), suggesting that both eigenvalues approach zero in the limit of large $j$.}
	\label{angleplot-2}
\end{figure}

\begin{figure}[b]
	\centering
		\includegraphics[width=0.65\textwidth]{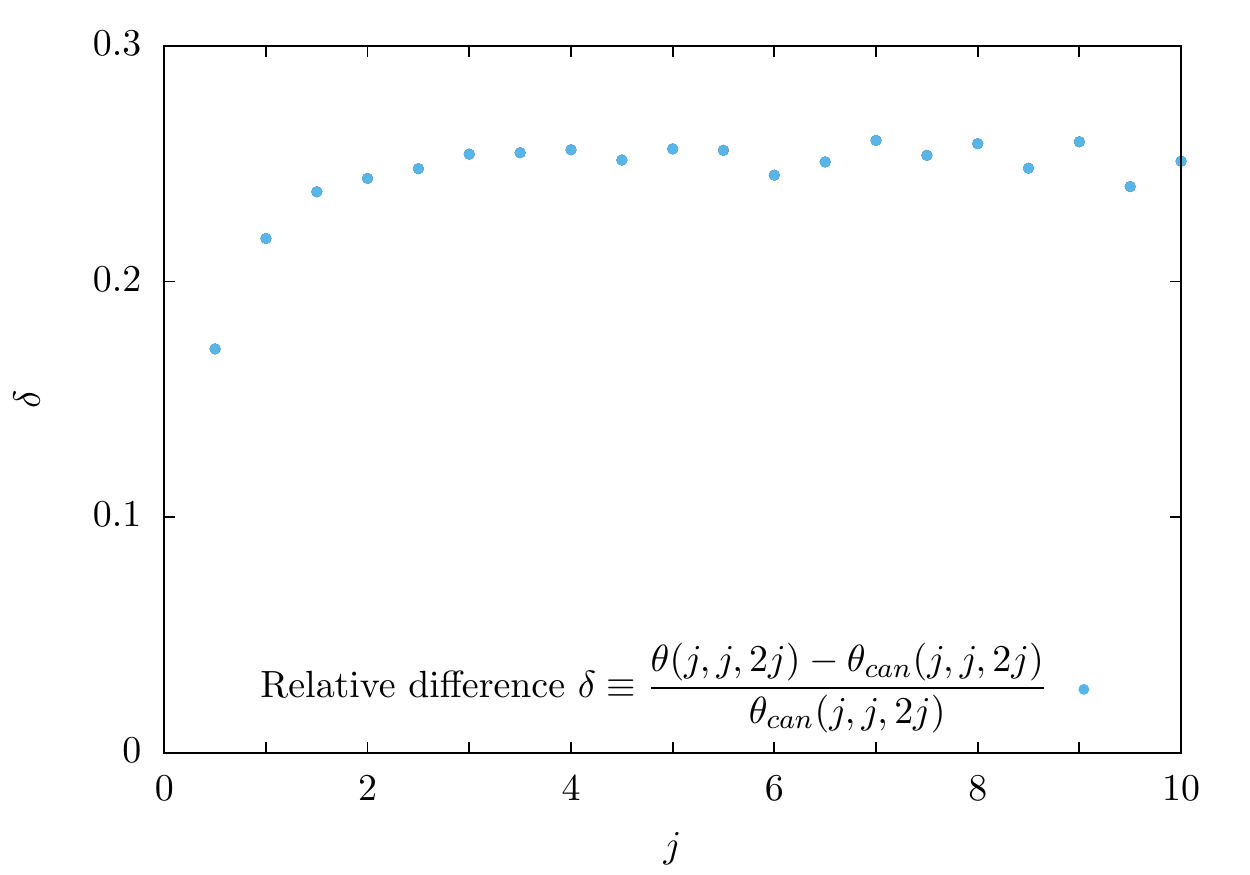}
		\caption{The relative difference between the eigenvalues of the canonical and coherent state angle operators in the degenerate case. As $j$ increases, $\delta$ remains roughly constant (being approximately equal to $25\,\%$).}
	\label{angle-diff}
\end{figure}

\clearpage }

\subsection{Volume operator}

The volume operator is associated to a node of the graph, and can be constructed as
\be\label{CS-volume}
\hat A_{V_n(\vec p_1,\dots,\vec p_N)} = \int d\mu(g_1,p_1)\cdots d\mu(g_N,p_N)\,V_n(\vec p_1,\dots,\vec p_N)\,\ket{g_1,p_1;\dots;g_N,p_N}\bra{g_1,p_1;\dots;g_N,p_N},
\ee
where $V_n(\vec p_1,\dots,\vec p_N)$ is a volume defined by the vectors $\vec p_1,\dots,\vec p_N$ associated with the $N$ links belonging to the node.\footnote
{
In the literature there exist several different proposals for the regularization of the classical volume, leading to different functions $V_n$ on the graph phase space. For this reason, we do not choose any specific such regularization, but keep $V_n$ unspecified and study the general properties of the corresponding coherent state operator. A more detailed analysis can be performed once a form for $V_n$ has been chosen: though an exact analysis (as in the case of the other operators introduced so far) is unlikely, we expect that saddle-point and numerical techniques can be used to estimate the spectrum and eigenstates of such operator.
}
The remark about choosing the $\vec p$-variables compatibly with the orientation of the graph naturally applies also here. We should assume that each $\vec p_l$ in Eq. \eqref{CS-volume} denotes either $\vec p_l$ or $\ppr_l$, depending on the orientation of the corresponding link.

The computation of the action of the operator \eqref{CS-volume} on spin network states is analogous to the corresponding calculation for the angle operator \eqref{CS-angle}. For this reason we refrain from showing the details of the calculation. For a three-valent node, one finds
\be
\hat A_{V_3}\Biggl|\makeSymbol{\includegraphics{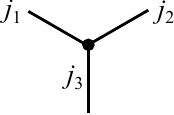}}\Biggr\rangle = v(j_1,j_2,j_3)\Biggl|\makeSymbol{\includegraphics{j1j2j3-node.pdf}}\Biggr\rangle,
\ee
where the eigenvalue is
\begin{align}\label{volume-3}
v(j_1,j_2,j_3) & = e^{-t(\lambda_{j_1}+\lambda_{j_2}+\lambda_{j_3})}\int d\nu(p_1)\,d\nu(p_2)\,d\nu(p_3)\,V_3(\vec p_1,\vec p_2,\vec p_3) \notag \\
&\times \begin{pmatrix} j_1&j_2&j_3 \\ m_1&m_2&m_3 \end{pmatrix} D^{(j_1)}_{m_1n_1}(e^{\vec p_1\cdot\vec\sigma}) D^{(j_2)}_{m_2n_2}(e^{\vec p_2\cdot\vec\sigma}) D^{(j_3)}_{m_3n_3}(e^{\vec p_3\cdot\vec\sigma}) \begin{pmatrix} j_1&j_2&j_3 \\ n_1&n_2&n_3 \end{pmatrix}.
\end{align}
Recalling the theorem of section \ref{positivity}, one should not expect the eigenvalue to be zero, unless the volume function $V_3(\vec p_1,\vec p_2,\vec p_3)$ is identically zero.

For a four-valent node, the matrix elements of the operator are given by
\begin{align}
\Biggl\langle \makeSymbol{\includegraphics{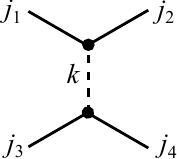}}\Biggr|\hat A_{V_4}\Biggl|\makeSymbol{\includegraphics{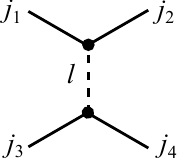}}\Biggr\rangle &= e^{-t(\lambda_{j_1}+\lambda_{j_2}+\lambda_{j_3}+\lambda_{j_4})}\int d\nu(p_1)\cdots d\nu(p_4)\,V(\vec p_1,\vec p_2,\vec p_3,\vec p_4) \notag \\
&{}\times \iota^{(k)}_{m_1m_2m_3m_4}D^{(j_1)}_{m_1n_1}(e^{\vec p_1\cdot\sigma})D^{(j_2)}_{m_2n_2}(e^{\vec p_2\cdot\sigma})D^{(j_3)}_{m_3n_3}(e^{\vec p_3\cdot\sigma})D^{(j_4)}_{m_4n_4}(e^{\vec p_4\cdot\sigma})\iota^{(l)}_{n_1n_2n_3n_4}. \label{volume-4}
\end{align}
The pattern displayed by Eqs. \eqref{volume-3} and \eqref{volume-4} generalizes to nodes of higher valence in an evident way.

The question of choosing the volume function $V_n(\vec p_1,\dots,\vec p_N)$ is not entirely clear, for two reasons. First, because there already exist three different proposals for the volume operator in LQG \cite{vol1, vol2, vol3}, differing in their diffeomorphism invariance properties and compatibility with a polyhedral description of spin networks. Any of these choices would in principle correspond to a different choice of the function $V_b$. 

Second, as it stands, the integration over the vectors $\vec p_1, \dots, \vec p_N$ in Eq. \eqref{CS-volume} is over all configurations of the vectors, rather than over closed configurations only. Consequently, the possible choices for $V_n$ are restricted to functions which either seem artificial, or are incompatible with the standard interpretation of the geometric content of a spin network node. Ideally, one would have a way of implementing the gauge invariance condition $\vec p_1 + \dots + \vec p_N = 0$ in the coherent state operator, allowing one to restrict the integration to closed configurations of the vectors. Then a preferred choice of $V_n$ would be the volume of the polyhedron spanned by the vectors $\vec p_1,\dots,\vec p_N$ summing up to zero. On the other hand, if one is willing to give up the picture of an $N$-valent node as an $N$-faced polyhedron, then one might wish to examine the consequences of choosing $V_n$ as the volume of the $(N+1)$-faced polyhedron spanned by $\vec p_1,\dots,\vec p_N$ with no condition of closure, implying in particular that three-valent nodes are carrying a non-zero volume.

\section{Conclusions} \label{conclu}

In this work, we presented a new proposal for constructing operators in LQG by using coherent states. The procedure allows one to uniquely associate an operator to a function on the classical phase space, and consists of inserting the classical function in the resolution of identity written in terms of coherent states. The resulting operators have a correct semiclassical limit by construction. As a first step, we introduced the coherent state operators corresponding to the basic canonical variables of LQG, i.e. the holonomy and the flux, and computed the (non-canonical) algebra of these operators. We also studied the elementary geometrical operators: area, angle and volume. Our computations show that these operators coincide with the canonical operators in the limit of large spins.

A positive feature of the operators we presented is that the operator corresponding to any classical function is immediately defined in an explicit way. This is in contrast with the situation of e.g. the canonical volume operator in LQG, where only the square of the volume operator is given by an explicit expression in terms of the fundamental operators, and to extract the volume operator itself one has to diagonalize a matrix whose dimension depends on the valence and spins of the node. It seems that nothing comes for free, though, since in our case the technical difficulties reappear in the form of integrals of functions depending on a large number of variables. However, if one is dealing with integrals instead of matrices, there is more hope of making progress through various analytical approximations, since several approximate techniques for computing integrals (such as the saddle point method) are available, and not as much is known about approximations related to diagonalizing large matrices. It is also conceivable that a different choice of coherent states as a starting point of our construction could lead to integrals which are easier to handle.

Our coherent state operators are by construction not cylindrically consistent, as they are based on functions on the phase space of a fixed graph. Nevertheless, we expect that once coherent states for the full theory (not restricted to a fixed graph) will be defined, it should be possible to employ such coherent states to define operators which would then be automatically cylindrically consistent.

For now, we consider the purpose of this work as a demonstration that operators alternative to the ones obtained through conventional canonical quantization can be constructed in the context of LQG. Whether they are better thought of as fundamental operators (alternative to the canonical ones), or as a technical tool providing a semiclassical approximation to the fundamental canonical operators, is not a question for the present work. As to the applications of our work, we hope that the procedure presented (of which we only gave examples for comparison with the canonical theory) can be used to define operators corresponding to classical functions for which a straightforward canonical quantization is problematic. In particular, we have in mind those classical functions which are not simple polynomials of the fundamental variables, such as the Hamiltonian constraint.

\section*{Acknowledgements}

This work was supported by the grant of Polish Narodowe Centrum
Nauki nr DEC-2011/02/A/ST2/00300 and nr 2013/09/N/ST2/04312. I.M. would like to thank the Jenny and Antti Wihuri Foundation for support.

\appendix

\section{$SU(2)$ recoupling theory}\label{intertwiners}

In this appendix we outline the elements of $SU(2)$ recoupling theory that we used to make the calculations in the main part of the paper. A more complete presentation of this material can be found e.g. in \cite{BrinkSatchler} and \cite{Varshalovich}. The formalism was first applied to LQG computations in \cite{recoupling1} using a calculus based on Temperley-Lieb algebras, and later refined in \cite{recoupling2,recoupling3, recoupling4}.

\subsection{Intertwiners}

The fundamental invariant tensor of $SU(2)$ is the epsilon tensor. In the spin-$j$ representation, it is given by
\be
\epsilon^{(j)}_{mn} = (-1)^{j-m}\delta_{m,-n}.
\ee
It satisfies the symmetry relation $\epsilon^{(j)}_{nm} = (-1)^{2j}\epsilon^{(j)}_{mn}$. The tensor $\epsilon^{(j)mn}$ is defined to be numerically equal to $\epsilon^{(j)}_{mn}$; then the contraction of two epsilons gives
\be
\epsilon^{(j)}_{m\mu}\epsilon^{(j)n\mu} = \delta_m^n.
\ee
Indices of $SU(2)$ tensors can be raised and lowered using the epsilon tensor as
\be
v^m = \epsilon^{mn}v_n, \qquad v_m = v^n\epsilon_{nm}.
\ee
The intertwiner between three representations $j_1$, $j_2$ and $j_3$ is given by the Wigner 3$j$-symbol:
\be
\iota_{m_1m_2m_3} = \begin{pmatrix} j_1&j_2&j_3 \\ m_1&m_2&m_3 \end{pmatrix}.
\ee
It is related to the Clebsch--Gordan coefficient by
\be\label{clebsch}
\begin{pmatrix} j_1&j_2&j_3 \\ m_1&m_2&m_3 \end{pmatrix} = \frac{(-1)^{j_1-j_2-m_3}}{\sqrt{d_{j_3}}}C^{j_1j_2j_3}_{m_1m_2\,-{}m_3}.
\ee
As an invariant tensor of $SU(2)$, the 3$j$-symbol satisfies
\be
D^{(j_1)}_{m_1n_1}(g)D^{(j_2)}_{m_2n_2}(g)D^{(j_3)}_{m_3n_3}(g)\begin{pmatrix} j_1&j_2&j_3 \\ n_1&n_2&n_3 \end{pmatrix} = \begin{pmatrix} j_1&j_2&j_3 \\ m_1&m_2&m_3 \end{pmatrix}
\ee
for any $g\in SU(2)$. By specializing to an infinitesimal transformation, one also has the relation
\be\label{3j-invariance}
{}^{(j_1)}\sigma^i_{m_1n_1}\begin{pmatrix} j_1&j_2&j_3 \\ n_1&m_2&m_3 \end{pmatrix} + {}^{(j_2)}\sigma^i_{m_2n_2}\begin{pmatrix} j_1&j_2&j_3 \\ m_1&n_2&m_3 \end{pmatrix} + {}^{(j_3)}\sigma^i_{m_3n_3}\begin{pmatrix} j_1&j_2&j_3 \\ m_1&m_2&n_3 \end{pmatrix} = 0.
\ee
Intertwiners of higher valence can be constructed by contracting several three-valent intertwiners. For example, a basis in the space of intertwiners between representations $j_1$, $j_2$, $j_3$ and $j_4$ is given by the objects
\be\label{iota4}
\iota^{(k)}_{m_1m_2;m_3m_4} = \iota_{m_1m_2n}\epsilon^{(k)nn'}\iota_{n'm_3m_4} = \begin{pmatrix} j_1&j_2&k \\ m_1&m_2&\mu \end{pmatrix}(-1)^{k-\mu}\begin{pmatrix} k&j_3&j_4 \\ -\mu&m_3&m_4 \end{pmatrix}.
\ee
Note that $\iota^{(k)}_{m_1m_2;m_3m_4}$ is not normalized; its norm is equal to $1/\sqrt{d_k}$.

\subsection{Graphical notation}

Calculations with intertwiners are conveniently made using a graphical notation, which we will now describe. The basic invariant tensors are represented graphically as follows:
\begin{align}
\delta^m_n &= \makeSymbol{\includegraphics[scale=1.25]{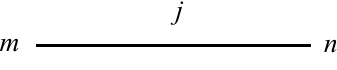}} \\
\epsilon^{(j)}_{mn} &= \makeSymbol{\includegraphics[scale=1.25]{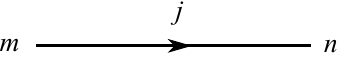}}
\end{align}
Contraction of magnetic indices is carried out simply by connecting the corresponding lines in the diagram. For example, the symmetry relation $\epsilon^{(j)}_{nm} = (-1)^{2j}\epsilon^{(j)}_{mn}$ and the contraction $\epsilon^{(j)}_{m\mu}\epsilon^{(j)n\mu} = \delta_m^n$ become
\begin{align}
\makeSymbol{\includegraphics[scale=1.25]{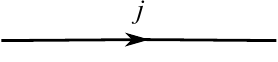}} &= (-1)^{2j}\makeSymbol{\includegraphics[scale=1.25]{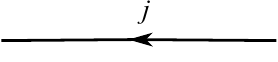}} \\
\makeSymbol{\includegraphics[scale=1.25]{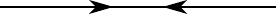}} &= \makeSymbol{\includegraphics[scale=1.25]{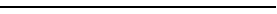}}
\end{align}
The 3$j$-symbol is represented by three lines connected at a node:
\be
\begin{pmatrix} j_1&j_2&j_3 \\ m_1&m_2&m_3 \end{pmatrix} = \makeSymbol{\includegraphics[scale=1.25]{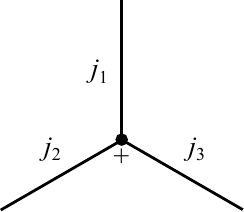}}
\ee
The order of spins in the symbol is encoded in a $+$ or $-$ at the node, corresponding respectively to counterclockwise and clockwise order. The 3$j$-symbol satisfies the symmetry relations
\begin{align}
\makeSymbol{\includegraphics[scale=1.25]{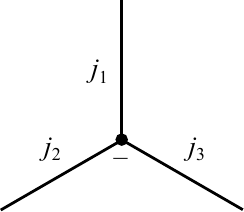}} &= (-1)^{j_1+j_2+j_3}\makeSymbol{\includegraphics[scale=1.25]{3j.pdf}} \\
\makeSymbol{\includegraphics[scale=1.25]{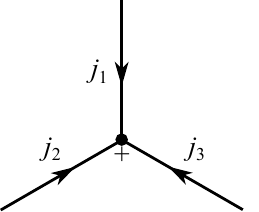}} &= \makeSymbol{\includegraphics[scale=1.25]{3j.pdf}}
\end{align}
and the orthogonality relation
\be
\makeSymbol{\includegraphics[scale=1.25]{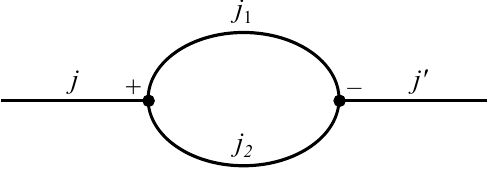}} = \delta_{jj'}\frac{1}{d_j}\;\makeSymbol{\includegraphics[scale=1.25]{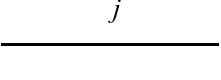}}
\ee
from which one also deduces the normalization
\be
\makeSymbol{\includegraphics[scale=1.25]{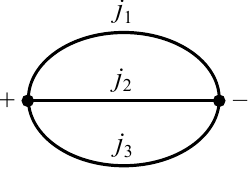}} = 1.
\ee
When one of the spins is zero, the 3$j$-symbol reduces to the epsilon tensor:
\be
\makeSymbol{\includegraphics[scale=1.25]{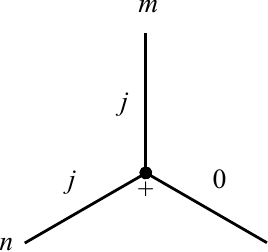}} = \frac{1}{\sqrt{d_j}}\;\makeSymbol{\includegraphics[scale=1.25]{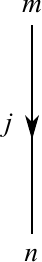}}
\ee
From Eq. \eqref{clebsch}, one deduces that the graphical representation of the Clebsch--Gordan coefficient is
\be\label{cg-picture}
C^{j_1j_2j}_{m_1m_2m} = (-1)^{j_1-j_2-j_3}\sqrt{d_j}\makeSymbol{\includegraphics[scale=1.25]{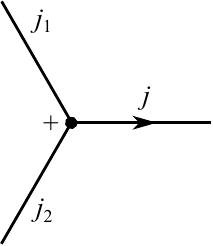}}
\ee
The $SU(2)$ generator\footnote{For the $j=1$ representation, one may consider the index $i$ to take values in the Cartesian basis (when $i=x,y,z$), or in the spherical basis (when $i=+,0,-$). The relation between the two bases is given by
\begin{align*}
v_+ = \frac{1}{\sqrt 2}(-v_x+iv_y), \qquad v_- = \frac{1}{\sqrt 2}(v_x+iv_y), \qquad v_0 = v_z; \\
v_x = \frac{1}{\sqrt 2}(-v_++v_-), \qquad v_y = -\frac{i}{\sqrt 2}(v_++v_-), \qquad v_z=v_0.
\end{align*}}
${}^{(j)}\sigma^i_{mn}$ is proportional to a Clebsch--Gordan coefficient; the precise relation is
\be\label{sigma-picture}
{}^{(j)}\sigma^i_{mn} = 2\sqrt{j(j+1)}C^{j1j}_{m1n} = 2\sqrt{j(j+1)(2j+1)} \makeSymbol{\includegraphics[scale=1.25]{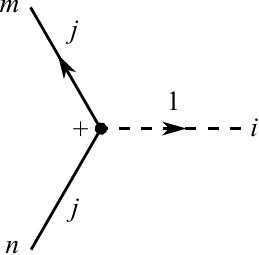}}
\ee
For completeness, we also give the graphical representation of the representation matrix of a $SU(2)$ element $g$:
\be\label{Dmatrix}
D^{(j)}_{mn}(g) = \makeSymbol{\includegraphics[scale=1.25]{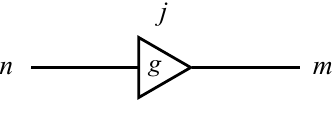}}
\ee

\subsection{The 6$j$-symbol}

A contraction of four 3$j$-symbols defines the 6$j$-symbol:
\be\label{6j-picture}
\begin{Bmatrix} j_1&j_2&j_3 \\ k_1&k_2&k_3 \end{Bmatrix} = \makeSymbol{\includegraphics[scale=1.25]{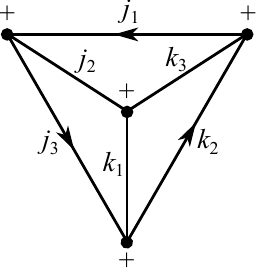}}
\ee
It appears for instance in the following relation, which is a special case of Eq. \eqref{4block} below, and which gives the change of basis between two different intertwiner bases of the form \eqref{iota4}:
\be
\makeSymbol{\includegraphics[scale=1.25]{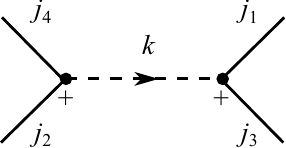}} = \sum_x d_x(-1)^{j_1+j_4-k-x}\begin{Bmatrix} j_1&j_2&x \\ j_4&j_3&k\end{Bmatrix} \makeSymbol{\includegraphics[scale=1.25]{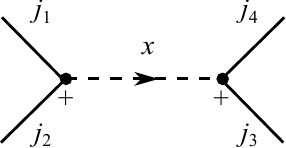}}
\ee
By performing the same change of basis in two steps, through an intermediate basis where $j_1$ is coupled to $j_3$, one deduces the relation
\be\label{6jsum}
\sum_x d_x(-1)^{x+k+l}\begin{Bmatrix} j_1&j_4&x \\ j_3&j_2&k \end{Bmatrix} \begin{Bmatrix} j_1&j_4&x \\ j_2&j_3&l \end{Bmatrix} = \begin{Bmatrix} j_1&j_2&k \\ j_4&j_3&l \end{Bmatrix}.
\ee
In calculations with the coherent state holonomy and flux operators, one also needs the following explicit expressions of 6$j$-symbols:
\be\label{6jzero}
\begin{Bmatrix} j_1&j_2&j_3 \\ j_2&j_1&0 \end{Bmatrix} = \frac{(-1)^{j_1+j_2+j_3}}{\sqrt{d_{j_1}d_{j_2}}},
\ee
\be\label{6jhalf}
\begin{Bmatrix} j&j&1 \\ \half&\half&j-\half \end{Bmatrix} = \frac{(-1)^{2j+1}}{\sqrt{6d_j}}\sqrt{\frac{j+1}{j}}, \qquad \begin{Bmatrix} j&j&1 \\ \half&\half&j+\half \end{Bmatrix} = \frac{(-1)^{2j+1}}{\sqrt{6d_j}}\sqrt{\frac{j}{j+1}}.
\ee

\subsection{Expanding invariant tensors in an intertwiner basis}\label{expanding}

An invariant tensor $t_{m_1\cdots m_N}$, having indices in representations $j_1,\dots,j_N$, is an element of the space of intertwiners between the representations $j_1,\dots,j_N$. As such, it can be expanded using a basis of the intertwiner space. Expressing an invariant tensor with $N$ indices as a block to which $N$ lines are attached, one has the relations
\begin{align}
\makeSymbol{\includegraphics[scale=1.25]{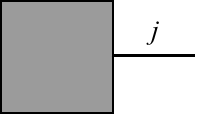}} &= \delta_{j,0}\;\makeSymbol{\includegraphics[scale=1.25]{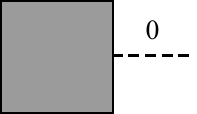}}\label{1block} \\
\makeSymbol{\includegraphics[scale=1.25]{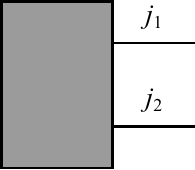}} &= \delta_{j_1j_2}\frac{1}{d_{j_1}}\;\makeSymbol{\includegraphics[scale=1.25]{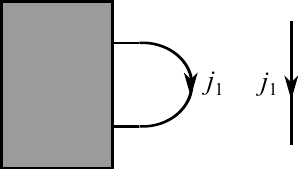}}\label{2block} \\
\makeSymbol{\includegraphics[scale=1.25]{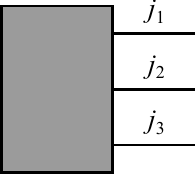}} &= \makeSymbol{\includegraphics[scale=1.25]{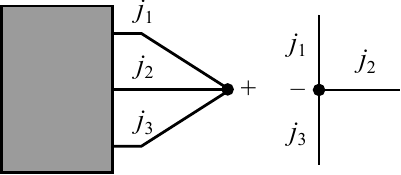}}\label{3block} \\
\makeSymbol{\includegraphics[scale=1.25]{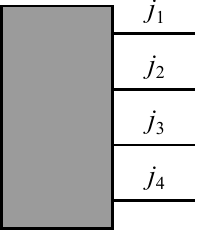}} &= \sum_x d_x\;\makeSymbol{\includegraphics[scale=1.25]{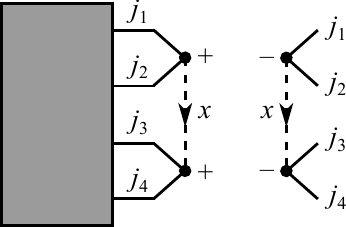}}\label{4block}
\end{align}
as well as the straightforward generalization of the last relation for tensors of higher order. The meaning of the first relation is that a tensor with a single free index will be invariant only if the index is in the trivial representation.

The method of expanding invariant tensors in intertwiners is very useful in evaluating certain kind of integrals over $SL(2, \mathbb{C})$, which we repeatedly encounter when evaluating matrix elements of coherent state operators between spin network states. Such integrals split into a part over $SU(2)$, and the remaining part. Typically, the $SU(2)$ part can be carried out using standard relations for integrals of Wigner matrices on $SU(2)$, while the remaining part can be reduced to the form
\be\label{oneintegral}
A_{mn} = \int d^3p\,f(\vec{p}) D^{(j)}_{mn}(e^{\alpha \vec{p} \cdot \vec{\sigma}}),
\ee
or, more generally,
\be\label{Nintegral}
A_{m_1\dots m_N;n_1\dots n_N} = \int d^3p_1\cdots d^3p_N\,f(\vec p_1,\dots,\vec p_N) D^{(j_1)}_{m_1n_1}(e^{\alpha \vec{p}_1 \cdot \vec{\sigma}})\cdots D^{(j_N)}_{m_Nn_N}(e^{\alpha \vec{p}_N \cdot \vec{\sigma}})
\ee
for some function $f$ and constant $\alpha$. 

The key observation for evaluating integrals such as \eqref{Nintegral} is that if the function $f$ is invariant under a common rotation of all its arguments, then the object $A_{m_1\dots m_N;n_1\dots n_N}$ is an invariant tensor of $SU(2)$, and as such it can be expanded in a basis of intertwiners. To prove invariance of $A_{m_1\dots m_N;n_1\dots n_N}$, we first note that
\be
D^{(j)}_{mm'}(g)D^{(j)}_{m'n'}(e^{\alpha\vec p\cdot\vec\sigma})D^{(j)}_{n'n}(g^\dagger) = D^{(j)}_{mn}(e^{\alpha\vec p^R\cdot\vec\sigma}),
\ee
where we have denoted the rotated vector as $(p^R)^i = R(g)^i_{\;j}p^j$, with $R(g)$ the $\mathbb{R}^3$ rotation matrix defined by the $SU(2)$ element $g$. We then find
\begin{align}
&D^{(j_1)}_{m_1m_1'}(g)\cdots D^{(j_N)}_{m_Nm_N'}(g) A_{m_1'\dots m_N';n_1'\dots n_N'} D^{(j_1)}_{n_1'n_1}(g^\dagger) \cdots D^{(j_N)}_{n_N'n_N}(g^\dagger) \notag \\
&= \int d^3p_1\cdots d^3p_N\,f(\vec p_1,\dots,\vec p_N) D^{(j_1)}_{m_1n_1}(e^{\alpha \vec{p}_1^R \cdot \vec{\sigma}})\cdots D^{(j_N)}_{m_Nn_N}(e^{\alpha \vec{p}_N^R\cdot \vec{\sigma}}) \notag \\
&= \int d^3p_1^R\cdots d^3p_N^R\,f(\vec p_1^R,\dots,\vec p_N^R) D^{(j_1)}_{m_1n_1}(e^{\alpha \vec{p}_1^R \cdot \vec{\sigma}})\cdots D^{(j_N)}_{m_Nn_N}(e^{\alpha \vec{p}_N^R \cdot \vec{\sigma}}) \notag \\
&= A_{m_1\dots m_N;n_1\dots n_N},
\end{align}
where we used rotational invariance of the function $f$ and the measure $d^3p$. From this we conclude that $A_{m_1\dots m_N;n_1\dots n_N}$ can be expanded in a basis of intertwiners as
\be \label{schur-general}
A_{m_1\dots m_N;n_1\dots n_N} = \sum_{\iota} c(\iota)\iota_{m_1\dots m_N;n_1\dots n_N}.
\ee
The coefficients in the expansion can be found by contracting each side of the equation with $\iota$; if the intertwiner basis is orthogonal, one finds $c(\iota) = (A\cdot\iota)/|\iota|^2$.

For the integral \eqref{oneintegral}, $SU(2)$ invariance is realized if $f(\vec p)$ depends only on the length of $\vec p$. In this case, the integral must be proportional to $\delta_{mn}$, which is the only invariant tensor with the correct index structure. Thus,
\be\label{schur-1}
A_{mn} = \int d^3p\,f(|\vec{p}|) D^{(j)}_{mn}(e^{\alpha \vec{p} \cdot \vec{\sigma}}) = c(j)\delta_{mn},
\ee
and by taking traces, the coefficient is found to be
\begin{align}
c(j) &= \dfrac{1}{d_j} \int d^3p \ f(|\vec{p}|) \chi^{(j)}(e^{\alpha \vec{p} \cdot \vec{\sigma}}) = \dfrac{1}{d_j} \int d^3p \ f(|\vec{p}|) \dfrac{\sinh(d_j\alpha|\vec p|)}{\sinh(\alpha|\vec p|)}.
\end{align}

\section{Clebsch-Gordan series for $e^{\vec p\cdot\vec\sigma/2}$}\label{cgseries-app}

In our calculations we also encounter products of Wigner matrices of the form $D^{(j)}_{mn}(e^{\vec p\cdot\vec\sigma/2})$. We wish to show that the standard Clebsch-Gordan series of $SU(2)$ can be extended to such products. We start by noting that, for a given element $e^{\vec{p} \cdot \vec{\sigma}/2}$, there always exists a rotation $R \in SU(2)$ such that
\be
R^\dag e^{\vec{p} \cdot \vec{\sigma}/2} R = e^{|\vec{p}| \sigma^3/2}.
\ee
(The physical interpretation of this statement is that a boost in an arbitrary direction can always be written as a boost along $z$ with respect to a rotated frame.) Using this, we can write
\begin{align}
D^{(j_1)}_{m_1 n_1}(e^{\vec{p} \cdot \vec{\sigma}/2}) & D^{(j_2)}_{m_2 n_2}(e^{\vec{p} \cdot \vec{\sigma}/2}) = D^{(j_1)}_{m_1 n_1}(RR^\dag e^{\vec{p} \cdot \vec{\sigma}/2} RR^\dag) D^{(j_2)}_{m_2 n_2}(RR^\dag e^{\vec{p} \cdot \vec{\sigma}/2} RR^\dag) \notag \\
&= \sum_{m_1'n_1'm_2'n_2'} D^{(j_1)}_{m_1 m'_1}(R) D^{(j_2)}_{m_2 m'_2}(R) D^{(j_1)}_{n'_1 n_1}(R^\dag) D^{(j_2)}_{n'_2 n_2}(R^\dag) D^{(j_1)}_{m'_1 n'_1}(e^{|\vec{p}| \sigma^3/2}) D^{(j_2)}_{m'_2 n'_2}(e^{|\vec{p}| \sigma^3/2}).
\end{align}
At this point, we use the standard Clebsch-Gordan series
\be\label{CGseries}
D^{(j_1)}_{m_1 n_1}(g) D^{(j_2)}_{m_2 n_2}(g) = \sum_{jmn} C^{j_1 j_2 j}_{m_1 m_2 m} C^{j_1 j_2 j}_{n_1 n_2 n} D^{(j)}_{mn}(g)
\ee
for a $g\in SU(2)$ to combine the two Wigner matrices in $R$ and the two in $R^\dag$. Inserting also $D^{(j)}_{mn}(e^{|\vec{p}| \sigma^3/2}) = \delta_{mn} e^{m |\vec{p}|}$, we get
\begin{align}
D^{(j_1)}_{m_1 n_1}(e^{\vec{p} \cdot \vec{\sigma}/2}) D^{(j_2)}_{m_2 n_2}(e^{\vec{p} \cdot \vec{\sigma}/2}) = & \sum_{\substack{jj'mn \\ m'n'm_1'm_2'}} C^{j_1 j_2 j}_{m_1 m_2 m} C^{j_1 j_2 j}_{m'_1 m'_2 n} C^{j_1 j_2 j'}_{m'_1 m'_2 m'} C^{j_1 j_2 j'}_{n_1 n_2 n'} D^{(j)}_{mn}(R) D^{(j')}_{m' n'}(R^\dag) e^{(m'_1 + m'_2) |\vec{p}|}
\end{align}
Now, $C^{j_1 j_2 j}_{m'_1 m'_2 n}$ is nonzero only if $m'_1 + m'_2 = n$, so we can replace $(m'_1+m'_2)$ with $n$ in the exponential. The sum over $m'_1$ and $ m'_2$ can then be carried out using an orthogonality relation of the Clebsch--Gordan coefficients. In the end we obtain
\be\label{newCGseries}
{D^{(j_1)}_{m_1 n_1}(e^{\vec{p} \cdot \vec{\sigma}/2})} D^{(j_2)}_{m_2 n_2}(e^{\vec{p} \cdot \vec{\sigma}/2}) = \sum_{jmn} C^{j_1 j_2 j}_{m_1 m_2 m} C^{j_1 j_2 j}_{n_1 n_2 n} D^{(j)}_{mn}(e^{\vec{p} \cdot \vec{\sigma}/2}),
\ee
which is what we were looking to prove.

\section{Operators constructed from gauge-invariant coherent states}\label{invstates}

We wish to compare the operators
\be
\hat A_f = \int d\mu(g_1,\vec p_1)\cdots d\mu(g_L,\vec p_L)\,f\bigl(\{g_l\},\{\vec p_l\}\bigr)\,\ket{\{g_l\}, \{\vec p_l\}}\bra{\{g_l\}, \{\vec p_l\}}
\ee
and
\be
\hat A_f^{\rm inv} = \int d\mu(g_1,\vec p_1)\cdots d\mu(g_L,\vec p_L)\,f\bigl(\{g_l\},\{\vec p_l\}\bigr)\,\ket{\Psi^t_{\Gamma,\{g_l\}, \{\vec p_l\}}}\bra{\Psi^t_{\Gamma,\{g_l\}, \{\vec p_l\}}},
\ee
which are constructed on a given graph $\Gamma$, and where $\ket{\{g_l\},\{\vec p_l\}}$ denotes the non-gauge invariant tensor product of single-link coherent states, while $\ket{\Psi^t_{\Gamma,\{g_l\}, \{\vec p_l\}}}$ is the gauge-invariant coherent state whose wave function is given in Eq. \eqref{inv-cs}.

We will show that $\hat A_f$ and $\hat A_f^{\rm inv}$ have the same action on spin network states based on the graph. To this end, we evaluate the matrix elements of both operators between two states of the form
\be
\ket{\Gamma,\{j_l\}, \{\iota_n\}} = \biggl(\prod_n \iota_n\biggr)^{n_1\cdots n_L}_{m_1\cdots m_L}\biggl(\prod_l \ket{j_l,m_l,n_l}\biggr).
\ee
We have
\begin{align}
&\bra{\Gamma,\{j_l\}, \{\iota_n\}} \hat A_f \ket{\Gamma,\{j_l'\},\{\iota_n'\}} \notag \\
&= \int d\mu(g_1,\vec p_1)\cdots d\mu(g_L,\vec p_L)\,f\bigl(\{g_l\},\{\vec p_l\}\bigr)\,\braket{\Gamma,\{j_l\}, \{\iota_n\}}{\{g_l\}, \{\vec p_l\}}\braket{\{g_l\}, \{\vec p_l\}}{\Gamma,\{j_l'\},\{\iota_n'\}}, \label{A-element}
\end{align}
and
\begin{align}
&\bra{\Gamma,\{j_l\}, \{\iota_n\}} \hat A_f^{\rm inv} \ket{\Gamma,\{j_l'\},\{\iota_n'\}}  \notag \\
&= \int d\mu(g_1,\vec p_1)\cdots d\mu(g_L,\vec p_L)\,f\bigl(\{g_l\},\{\vec p_l\}\bigr)\,\braket{\Gamma,\{j_l\}, \{\iota_n\}}{\Psi^t_{\Gamma,\{g_l\}, \{\vec p_l\}}}\braket{\Psi^t_{\Gamma,\{g_l\}, \{\vec p_l\}}}{\Gamma,\{j_l'\},\{\iota_n'\}}.\label{Ainv-element}
\end{align}
From Eq. \eqref{inv-cs}, we know that
\be
\braket{\Gamma,\{j_l\}, \{\iota_n\}}{\Psi^t_{\Gamma,\{j_l\}, \{\iota_n\}}}  = e^{-t(\lambda_{j_1} + \dots + \lambda_{j_L})/2} \overline{\Phi_{\Gamma,\{j_l\}, \{\iota_n\}}(\{g_l e^{t\vec p_l\cdot\vec\sigma/2}\})}.
\ee
On the other hand, in the group representation we can compute
\begin{align}
&\braket{\Gamma,\{j_l\}, \{\iota_n\}}{\{g_l\}, \{\vec p_l\}} \notag \\
&= \int d\tilde g_1\cdots d\tilde g_L\,\biggl(\prod_n \iota_n\biggr)^{n_1\cdots n_L}_{m_1\cdots m_L}\biggl(\prod_l\sqrt{d_{j_l}}D^{(j_l)}_{m_ln_l}(\tilde g_l)\biggr) \biggl(\prod_l d_{j_l}e^{-t\lambda_{j_l}/2}\chi^{(j_l)}(g_le^{t\vec p_l\cdot\vec\sigma/2}\tilde g_l^{-1})\biggr) \notag \\
&= \biggl(\prod_n \iota_n\biggr)^{n_1\cdots n_L}_{m_1\cdots m_L}\biggl(\prod_l \sqrt{d_{j_l}}e^{-t\lambda_{j_l}/2}D^{(j_l)}_{m_ln_l}(g_le^{t\vec p_l\cdot\vec\sigma/2})\biggr)\notag \\
&= e^{-t(\lambda_{j_1} + \dots + \lambda_{j_L})/2}\overline{\Phi_{\Gamma,\{j_l\}, \{\iota_n\}}(\{g_l e^{t\vec p_l\cdot\vec\sigma/2}\})}.
\end{align}
This shows that the right-hand sides of Eqs. \eqref{A-element} and \eqref{Ainv-element} are equal.

\end{document}